\documentclass[11pt,a4paper]{article}
\pdfoutput=1

\usepackage{bmpsize}
\usepackage{jheppub}

\usepackage{graphicx,bm,color,wrapfig,cancel}
\usepackage{amsmath,amssymb,bm}
\usepackage{float}
\usepackage{braket}
\usepackage{comment}
\usepackage{subfigure}
\usepackage[utf8]{inputenc}
\usepackage{appendix}
\usepackage{ulem}

\usepackage{mathrsfs}

\usepackage{here}
\usepackage{tikz}
\usetikzlibrary{matrix}

\newcommand{\Slash}[1]{{\ooalign{\hfil/\hfil\crcr$#1$}}}

\newcommand{\fulltoday}{\number\day\space \ifcase\month\or
    January\or February\or March\or April\or May\or June\or
    July\or August\or September\or October\or November\or December\fi
    \space\number\year}

 \makeatletter
    
    \@addtoreset{equation}{section}
  \makeatother

\allowdisplaybreaks[3]

\title{\boldmath
Fate of the topological susceptibility in two-color dense QCD 
}

\author[a]{Mamiya Kawaguchi,}
\author[b,c]{Daiki~Suenaga}

\affiliation[a]{School of Nuclear Science and Technology, University of Chinese Academy of Sciences, Beijing 100049, China}

\affiliation[b]{
Few-Body Systems in Physics Laboratory, RIKEN Nishina Center, Wako 351-0198, Japan}

\affiliation[c]{
Research Center for Nuclear Physics,
Osaka University, Ibaraki 567-0048, Japan }

\emailAdd{mamiya@ucas.ac.cn}
\emailAdd{daiki.suenaga@riken.jp}

\abstract{
We explore the topological susceptibility at finite quark chemical potential and zero temperature in two-color QCD (QC$_2$D) with two flavors. Through the Ward-Takahashi identities of QC$_2$D, we find that the topological susceptibility in the vacuum solely depends on three observables: the pion decay constant, the pion mass, and the $\eta$ mass in the low-energy regime of QC$_2$D. Based on the identities, we numerically evaluate the topological susceptibility at finite quark chemical potential using the linear sigma model with the approximate Pauli-Gursey $SU(4)$ symmetry. Our findings indicate that, in the absence of $U(1)_A$ anomaly effects represented by the Kobayashi-Maskawa-'t Hooft-type determinant interaction, the topological susceptibility vanishes in both the hadronic and baryon superfluid phases. On the other hand, when the $U(1)_A$ anomaly effects are present, the constant and nonzero topological susceptibility is induced in the hadronic phase, reflecting the mass difference between the pion and $\eta$ meson. Meanwhile, in the superfluid phase it begins to decrease smoothly. The asymptotic behavior of the decrement is fitted by the continuous reduction of the chiral condensate in dense QC$_2$D, which is similar to the behavior observed in hot three-color QCD matter. In addition, effects from the finite diquark source on the topological susceptibility are discussed. We expect that the present study provides a clue to shed light on the role of the $U(1)_A$ anomaly in cold and dense QCD matter.
}

\begin{document}

\maketitle
\flushbottom




\section{Introduction}

In Quantum Chromodynamics (QCD), the $U(1)_A$ anomaly, i.e., the non-conservation of the $U(1)_A$ axial current caused by the gluonic quantum corrections, plays crucial roles in the low-energy physics governed by the spontaneous breaking of the chiral symmetry. For instance, the $U(1)_A$ anomaly affects the hadron mass spectrum to yield the heavy $\eta'$ meson~\cite{Weinberg:1975ui} and
the order of the chiral phase transition in QCD matter~\cite{Buballa:2003qv}. In addition to these low-energy aspects, the $U(1)_A$ anomaly is also closely related with topological vacuum structures of QCD~\cite{tHooft:1986ooh}, which is described by the anomalous gluonic operator tagged with the $\theta$ parameter. The characteristics of the $\theta$-dependent QCD vacuum is captured by the {\it topological susceptibility}: the curvature of the QCD effective potential with respect to $\theta$.

The symmetry breakings in QCD are reflected in meson susceptibility functions defined by two-point functions of quark composite operators in the low-energy limit. At the hadronic level, the so-called {\it chiral-partner structure} would be an indication of this property~\cite{Hatsuda:1994pi}. That is, at high temperature and/or density where chiral symmetry tends to be restored, masses of the mesons related by the chiral transformation become degenerate, and so do the corresponding meson susceptibility functions. This implies that, indeed, the meson susceptibility functions can be regarded as alternative probes to measure the strength of chiral symmetry breaking and restoration. In a similar way, the effective restoration of $U(1)_A$ symmetry can be quantified by the degeneracies of the meson susceptibility functions connected by the $U(1)_A$ transformation.

Making use of the Ward-Takahashi identity (WTI) associated with chiral symmetry, one can show that the topological susceptibility is also correlated with the chiral- and $U(1)_A$-partner structures in the meson susceptibility functions~\cite{GomezNicola:2016ssy,GomezNicola:2017bhm,Kawaguchi:2020qvg,Cui:2021bqf,Cui:2022vsr}. Thereby, the topological susceptibility can also be referred to as the indicator for the breaking strength of $U(1)_A$ symmetry through the chiral phase transition. 
In fact, lattice QCD simulations at the physical quark masses support that the magnitude of the topological susceptibility smoothly decreases at high temperatures~\cite{Petreczky:2016vrs,Borsanyi:2016ksw,Bonati:2018blm}.
Furthermore, 
a strong correlation with the chiral restoration has also been studied through the meson susceptibility functions within $2$ flavor QCD~\cite{Cohen:1996ng,Aoki:2012yj,Tomiya:2016jwr,Aoki:2020noz,Suzuki:2020rla}
and $2+1$ flavor QCD~\cite{HotQCD:2012vvd,Buchoff:2013nra,Bhattacharya:2014ara}.


Thus far, the susceptibilities in hot QCD matter have been explored by both lattice simulations~\cite{Cohen:1996ng,Aoki:2012yj,HotQCD:2012vvd,Buchoff:2013nra,Bhattacharya:2014ara,Tomiya:2016jwr,Petreczky:2016vrs,Borsanyi:2016ksw,Bonati:2018blm,Aoki:2020noz,Suzuki:2020rla} and effective model analyses~\cite{Kawaguchi:2020qvg,Kawaguchi:2020kdl,Cui:2021bqf}
in order to gain deeper insights into the symmetry properties of QCD in the extreme environment. However, at finite quark chemical potential $\mu_q$ lattice QCD simulations with three colors suffer from the {\it sign problem}, and then the first-principle numerical computations cannot apply in baryonic matter straightforwardly~\cite{Aarts:2015tyj}. For this reason, our understanding of QCD at low-temperature and high-density regime is still limited compared to that in hot medium.

In light of the difficulty of three-color QCD on lattice simulations with finite $\mu_q$, two-color QCD (${\rm QC_2D}$) with two flavors provides us with a valuable testing ground. This is because the sign problem is resolved in such QCD-like theory owing to its pseudo-real property~\cite{Muroya:2003qs}. Focusing on this fact, many efforts from lattice simulations are being devoted to understandings of, e.g., phase structures, thermodynamics quantities, electromagnetic responses, the hadron mass spectrum, and gluon propagators in cold and dense QC$_2$D matter~\cite{Hands:1999md,Kogut:2001na,Hands:2001ee,Muroya:2002ry,Chandrasekharan:2006tz,Hands:2006ve,Alles:2006ea,Hands:2007uc,Hands:2010gd,Hands:2011hd,Cotter:2012mb,Hands:2012yy,Boz:2013rca,Braguta:2016cpw,Puhr:2016kzp,Boz:2018crd,Astrakhantsev:2018uzd,Iida:2019rah,Wilhelm:2019fvp,Buividovich:2020gnl,Iida:2020emi,Astrakhantsev:2020tdl,Bornyakov:2020kyz,Buividovich:2020dks,Buividovich:2021fsa,Iida:2022hyy}. In association with such numerical examinations, theoretical investigations of QC$_2$D at finite $\mu_q$ based on effective models have been done~\cite{Kogut:1999iv,Kogut:2000ek,Lenaghan:2001sd,Ratti:2004ra,Sun:2007fc,Brauner:2009gu,Kanazawa:2009ks,Harada:2010vy,Strodthoff:2011tz,Suenaga:2019jjv,Contant:2019lwf,Khunjua:2020xws,Kojo:2021knn,Suenaga:2021bjz,Kojo:2021hqh,Suenaga:2022uqn}.

In QC$_2$D, diquarks composed of two quarks are treated as color-singlet baryons. In other words, baryons exhibit bosonic behavior similarly to mesons. Reflecting this fact in QC$_2$D, $SU(2)_L\times SU(2)_R$ chiral symmetry is extended to the so-called Pauli-Gursey $SU(4)$ symmetry, which allows us to describe diquark baryons and light mesons in the single multiplets. Accordingly, the spontaneously symmetry-breaking pattern caused by the chiral condensate is changed to $SU(4)\to Sp(4)$~\cite{Kogut:1999iv,Kogut:2000ek}. Despite such an extension of chiral symmetry, symmetry structures of the $U(1)_A$ axial anomaly induced by gluonic configurations essentially do not differ from those in ordinary three-color QCD. 


To shed light on the role of the $U(1)_A$ axial anomaly in baryonic QCD matter, 
the $\mu_q$ dependence of the topological susceptibility has been numerically measured by lattice numerical simulations of QC$_2$D~\cite{Alles:2006ea,Hands:2011hd,Iida:2019rah,Astrakhantsev:2020tdl}.
The recent lattice result in Ref.~\cite{Iida:2019rah} indicates that 
the effect of $\mu_q$ does not exert any influence on the behavior of the topological susceptibility in baryonic matter, resulting in an approximately constant value. In contrast, the other group shows that the topological susceptibility is suppressed in high-density regions~\cite{Astrakhantsev:2020tdl}. Hence, there exist discrepancies among the lattice simulations at finite quark chemical potentials, and the fate of topological susceptibility at high-density regions is still controversial. 

In this paper, motivated by the above puzzle, we investigate the topological susceptibility in zero-temperature QC$_2$D at finite $\mu_q$ based on an effective-model approach. In particular, we employ the linear sigma model based on the approximate Pauli-Gursey $SU(4)$ symmetry invented in Ref.~\cite{Suenaga:2022uqn}. Notably, this model is capable of treating the $\eta$ meson which plays a significant role in describing the $U(1)_A$ anomaly structures consistently with other light mesons and diquark baryons. In QC$_2$D, since the diquarks obey the Bose-Einstein statistics, when the mass of the ground-state diquark becomes zero they begin to exhibit the Bose-Einstein condensates (BECs), leading to the emergence of the {\it diquark condensed phase}~\cite{Kogut:1999iv,Kogut:2000ek}. This phase is also referred to as the {\it baryon superfluid phase} due to the violation of $U(1)_B$ baryon-number symmetry. Meanwhile, the stable phase with no such BECs connected to the vacuum, i.e., zero temperature and zero chemical potential, is called the {\it hadronic phase}. The former nontrivial phase triggers a rich hadron mass spectrum such as a mixing among hadrons sharing the identical quantum numbers except for the baryon number. 


Within the linear sigma model, the influences of the $U(1)_A$ anomaly on hadrons are described by the so-called Kobayashi-Maskawa-'t Hooft (KMT)-type determinant interaction~\cite{Kobayashi:1970ji,Kobayashi:1971qz,tHooft:1976snw,tHooft:1976rip}, which only breaks $U(1)_A$ symmetry but preserves the Pauli-Gursey $SU(4)$ one. This interaction induces a mass difference between the pion and $\eta$ meson in the vacuum. Thus, in the present analysis we particularly focus on the strength of the KMT-type interaction, in other words, the mass difference between the pion and $\eta$ meson, in order to quantify roles of the $U(1)_A$ anomaly in the topological susceptibility. Besides, in lattice simulations source contributions with respect to the diquark condensate would be left sizable, so in this paper we also investigate the diquark source effects so as to facilitate the comparison with lattice data.

This paper is organized as follows. In Sec.~\ref{sec:chitop_derivation} we present general properties associated with the topological susceptibility in QC$_2$D by focusing on the underlying QC$_2$D theory, and discuss symmetry partner structures of the meson susceptibility functions. In Sec.~\ref{sec:LSM}, the emergence of the Pauli-Gursey $SU(4)$ symmetry in QC$_2$D is briefly explained, and our linear sigma model regarded as a low-energy theory of QC$_2$D is introduced. In Sec.~\ref{sec:Analysis} we show how the topological susceptibility within the linear sigma model is evaluated by explicitly demonstrating the matching between underlying QC$_2$D and the linear sigma model. Based on it, in Sec.~\ref{sec:Numerical} we show our numerical results on the topological susceptibility at finite $\mu_q$. In order to facilitate the comparison with lattice simulations, in Sec.~\ref{sec:FiniteJ} we also exhibit the results in the presence of the diquark source contributions. Finally, in Sec.~\ref{sec:Conclusion} we conclude our present study.

\section{
Topological susceptibility based on Ward-Takahashi identities of QC$_2$D
}
\label{sec:chitop_derivation}

Our main aim of this paper is to reveal properties of the topological susceptibility in zero-temperature QC$_2$D with finite quark chemical potential $\mu_q$. 
In this section, we present an analytic formula of the topological susceptibility based on the underlying QC$_2$D Lagrangian~\cite{GomezNicola:2016ssy,GomezNicola:2017bhm,Kawaguchi:2020qvg,Cui:2021bqf,Cui:2022vsr}, which is useful for the investigation within the effective-model framework of the linear sigma model.

The topological susceptibility is one of indicators to measure the magnitude of the $U(1)_A$ anomaly, which is related to nontrivial gluonic configurations such as the instantons~\cite{tHooft:1986ooh}. Hence, we need to return to QCD Lagrangian where such microscopic degrees of freedom are treated manifestly. In two-flavor QC$_2$D, the Lagrangian including the so-called QCD $\theta$-term in Minkowski spacetime is of the form 
\begin{eqnarray}
{\cal L}_{\rm QC_2D}=
\bar \psi (i\gamma^\mu D_\mu-m_l)\psi - \frac{1}{4}G_{\mu\nu}^aG^{\mu\nu,a}
+\theta\frac{g^2}{64\pi^2}\epsilon^{\mu\nu\rho\sigma} G_{\mu\nu}^aG_{\rho\sigma}^a \ .
\label{QCD_lag_theta}
\end{eqnarray}
As for the first term, $\psi=(u,d)^T$ denotes the two-flavor quark doublet and $D_\mu\psi=(\partial_\mu-i\mu_q\delta_{\mu0}-igA_\mu^aT_c^a)\psi$ is the covariant derivative incorporating effects from a quark chemical potential $\mu_q$ and interactions with a gluon field $A_\mu^a$. The $2\times2$ matrix $T_c^a=\tau_c^a/2$ is the generator of $SU(2)_c$ color group with $\tau_c^a$ being the Pauli matrix. Besides, $g$ and $m_l$ are the QCD coupling constant and a current quark mass where the isospin symmetric limit is taken, $m_u=m_d\equiv m_l$. 
The second term in Eq.~(\ref{QCD_lag_theta}) is a gluon kinetic term where $G_{\mu\nu}^a=\partial_\mu A_\nu^a-\partial_\nu A_\mu^a+g\epsilon^{abc}A_\mu^b A_\nu^b$ is the field strength of gluons. The last ingredient of ${\rm QC_2D}$ Lagrangian in Eq.~(\ref{QCD_lag_theta}) is 
the $\theta$-term of QC$_2$D, which is described by a flavor-singlet topological operator $Q\equiv(g^2/64\pi^2)\epsilon^{\mu\nu\rho\sigma} G_{\mu\nu}^aG_{\rho\sigma}^a$ tagged with the $\theta$-parameter. Our purpose in this subsection is to derive useful identities with respect to the topological susceptibility, so that only the $\theta$-dependent term in Eq.~(\ref{QCD_lag_theta}), which is gauge invariant, plays significant roles. For this reason, the gauge-fixing terms and the corresponding Faddeev-Popov determinant, which do not affect the following discussions, have been omitted in Eq.~(\ref{QCD_lag_theta}).

The generating functional of ${\cal L}_{\rm QC_2D}$ in the path-integral formulation is given by
\begin{eqnarray}
Z_{\rm QC_2D}
=
\int [d\bar{\psi} d\psi][dA]
\exp\Biggl[
i\int d^4x
{\cal L}_{\rm QC_2D}
\Biggl]\ , \label{ZQC2D}
\end{eqnarray}
and the $\theta$-dependent 
effective action 
of ${\rm QC_2D}$ 
is evaluated as  
\begin{eqnarray}
\Gamma_{{\rm QC_2D}} =-i \ln Z_{{\rm QC_2D} }\ . \label{VQC2D}
\end{eqnarray}
The topological susceptibility $\chi_{\rm top}$ is defined by the curvature of $\Gamma_{{\rm QC_2D}}$, i.e., a second derivative with respect to $\theta$ at $\theta=0$:
\begin{eqnarray}
\chi_{\rm top}&=&
-\int d^4x\frac{\delta^2 \Gamma_{{\rm QC_2D}}}{\delta \theta (x) \delta \theta (0)}\Biggl|_{\theta=0} .
\label{def_chitop}
\end{eqnarray}
Thus, from a straightforward calculation of Eq.~(\ref{def_chitop}) based on the QC$_2$D Lagrangian in Eq.~(\ref{QCD_lag_theta}), one can find that the topological susceptibility is described by a two-point correlation function of 
the topological operator $Q=(g^2/64\pi^2)\epsilon^{\mu\nu\rho\sigma} G_{\mu\nu}^aG_{\rho\sigma}^a$:
\begin{eqnarray}
\chi_{\rm top}
&=&
-i\int d^4x \langle0|TQ(x) Q(0) |0\rangle\ ,
\label{chitop_Q}
\end{eqnarray}
with $T$ denoting the time ordered product. It should be noted that contributions stemming from a product of $\langle Q\rangle$ have been omitted from Eq.~(\ref{chitop_Q}) due to the parity conservation. The topological susceptibility in Eq.~(\ref{chitop_Q}) is written in terms of the gluonic operator $Q$, which would not be a manageable expression since our task in this paper is to evaluate $\chi_{\rm top}$ from a low-energy effective model involving only hadronic degrees of freedom. Difficulties in matching the susceptibility from the effective models with that from underlying QC$_2$D are, however, remedied by utilizing the $U(1)_A$ axial rotation properly as demonstrated below.

Under the $U(1)_A$ rotation with a rotation angle $\alpha_A$, the quark doublet transforms as
\begin{eqnarray}
\psi\to \exp(i\alpha_A/2\, \gamma_5)\psi\ .
\end{eqnarray}
Meanwhile, within the path-integral formalism the gluonic quantum anomaly is generated by the fermionic measure $[d\bar{\psi}d\psi]$ according to the Fujikawa's method~\cite{Fujikawa:1979ay}, resulting in that the rotated generating functional reads
\begin{eqnarray}
Z_{\rm QC_2D}
&\to&
\int [d\bar{\psi} d\psi][dA]
\exp\Biggl[
i\int d^4x
\Biggl(
\bar \psi i\gamma^\mu D_\mu \psi 
-m_l\bar \psi \exp(i\alpha_A \gamma_5)\psi
\nonumber\\
&&
-\frac{1}{4}G_{\mu\nu}^aG^{\mu\nu, a}
+(\theta-2\alpha_A)\frac{g^2}{64\pi^2}\epsilon^{\mu\nu\rho\sigma} G_{\mu\nu}^aG_{\rho\sigma}^a
\Biggl)
\Biggl]\ . \label{ZAxialTrans}
\end{eqnarray}
Thus, when choosing the rotation angle to be $\alpha_A=\theta/2$, the $\theta$-dependence of the QCD $\theta$-term is transferred into the quark mass term as
\begin{eqnarray}
Z_{\rm QC_2D}&=&
\int [d\bar{\psi} d\psi][dA]
\exp\Biggl[
i\int d^4x
\Biggl(
\bar \psi i\gamma^\mu D_\mu \psi 
-m_l\bar \psi \exp\left(i\theta/2\, \gamma_5\right)\psi
-\frac{1}{4}G_{\mu\nu}^aG^{\mu\nu, a}
\Biggl)
\Biggl]\ . \nonumber\\
\label{gene_func_mass_theta}
\end{eqnarray}
Following the procedure in Eqs.~(\ref{VQC2D}) and~(\ref{def_chitop}) with the rotated generating functional~(\ref{gene_func_mass_theta}), the topological susceptibility $\chi_{\rm top}$ is now expressed by fermionic operators as
\begin{eqnarray}
\chi_{\rm top}=
-\frac{1}{4}
\left[
m_l\langle \bar \psi \psi\rangle +im_l^2 \chi_{\eta} 
\right]\ ,
\label{chitop_quark}
\end{eqnarray}
where $\langle\bar{\psi}\psi\rangle$ is the chiral condensate serving as an order parameter of the spontaneous chiral-symmetry breaking, and $\chi_{\eta}$ denotes an $\eta$-meson susceptibility function defined by
\begin{eqnarray}
\chi_{\eta}=\int d^4x \langle 0|T (i\bar \psi\gamma_5 \psi)(x) (i\bar \psi\gamma_5 \psi)(0)   |0\rangle\ .
\label{eta_sus_QCD}
\end{eqnarray}

It should be noted that, from the rotated generating functional in Eq.~(\ref{ZAxialTrans}),
a non-conservation law of the $U(1)_A$ axial current $j^\mu_A=\bar{\psi}\gamma^\mu\gamma_5\psi$ is also obtained as
\begin{eqnarray}
\partial_\mu j_A^\mu = 2m_l\bar{\psi}i\gamma_5\psi + \frac{g^2}{16\pi^2}\epsilon^{\mu\nu\rho\sigma}G_{\mu\nu}^a G_{\rho\sigma}^a\ . \label{AxialD}
\end{eqnarray}

The topological susceptibility~(\ref{chitop_quark}) is further reduced to a handleable form. In fact, using the $SU(2)_L\times SU(2)_R$ chiral-partner relation shown in Appendix~\ref{sec:Trasformation}, a chiral WTI with respect to the chiral condensate $\langle\bar{\psi}\psi\rangle$ is derived as in Appendix~\ref{sec:WTIs}, which reads 
\begin{eqnarray}
\langle \bar \psi \psi\rangle=-im_l\chi_{\pi}\ . 
\label{AWI_chiral}
\end{eqnarray}
In this identity, $\chi_\pi$ is a pion susceptibility function defined by
\begin{eqnarray}
\chi_{\pi}\delta^{ab}=
\int d^4x \langle 0|T (i\bar \psi\gamma_5 \tau^a_f \psi)(x) (i\bar \psi\gamma_5 \tau^b_f \psi)(0)   |0\rangle\ , \label{PiSusQCD}
\end{eqnarray}
with $\tau^a_f$ being the Pauli matrix in the flavor space. Therefore, inserting Eq.~(\ref{AWI_chiral}) into Eq.~(\ref{chitop_quark}), 
the topological susceptibility is found to be determined in terms of a difference of $\chi_{\pi} $ and $\chi_{\eta}$ as
\begin{eqnarray}
\chi_{\rm top}=
\frac{im_l^2}{4}
(
\chi_{\pi} - \chi_{\eta}
)\ .
\label{chitop_meson_sus}
\end{eqnarray}
This expression is identical to the one obtained in ordinary three-color QCD through the WTI~\cite{GomezNicola:2016ssy,GomezNicola:2017bhm,Kawaguchi:2020qvg,Cui:2021bqf,Cui:2022vsr}.
Here, to facilitate an understanding of the role of topological susceptibility, 
we insert the scalar meson susceptibilities $\chi_{\sigma}$ and $\chi_{a_0}$ in Eq.~(\ref{chitop_meson_sus}):
\begin{eqnarray}
\chi_{\rm top}&=&
\frac{im_l^2}{4}
\left[(\chi_{\pi}-\chi_\sigma) - (\chi_{\eta}-\chi_\sigma)
\right]\ ,\nonumber\\ 
\chi_{\rm top}&=&
\frac{im_l^2}{4}
\left[(\chi_{\pi}-\chi_{a_0}) - (\chi_{\eta}-\chi_{a_0})
\right]\ ,
\label{chitop_facilitate }
\end{eqnarray}
where $\chi_\sigma$ and $\chi_{a_0}$ are the susceptibility functions made of the composite operators $\bar{\psi}\psi$ and $\bar{\psi}\tau_f^a\psi$, respectively. Indeed,
under the chiral $SU(2)_L\times SU(2)_R$ rotation and the $U(1)_A$ rotation,
the meson susceptibility functions are transformed into each other:
\begin{center}
\begin{tikzpicture}
  \matrix (m) [matrix of math nodes,row sep=4em,column sep=5em,minimum width=3em]
  {
     \chi_{\pi} & \chi_{\sigma} \\
     \chi_{a_0} & \chi_{\eta} \\};
  \path[-stealth]
    (m-1-1) edge node [midway,left] {$U(1)_A$} (m-2-1)
            edge node [above] {SU(2)} (m-1-2)
    (m-2-1) edge node [below] {SU(2)} (m-2-2)
    (m-1-2) edge node [right] {$U(1)_A$} (m-2-2)
    (m-2-1) edge node [midway,left] { } (m-1-1)
    (m-1-2) edge node [midway,left] { } (m-1-1)
    (m-2-2) edge node [midway,left] { } (m-1-2)
    (m-2-2) edge node [midway,left] { } (m-2-1);
\end{tikzpicture}
\end{center}
as explicitly shown in Appendix~\ref{sec:Trasformation}.
With this transformation, one can realize that 
the topological susceptibility in Eq.~(\ref{chitop_facilitate }) is described by
the combinations of the chiral $SU(2)$ partner $\chi_\pi \leftrightarrow \chi_\sigma$ ($\chi_{a_0} \leftrightarrow \chi_\eta$) and the $U(1)$ axial partner $\chi_\pi \leftrightarrow \chi_{a_0}$ ($\chi_{\sigma} \leftrightarrow \chi_\eta$) . 
When chiral symmetry is restored and the order parameter of the spontaneous chiral symmetry breaking vanishes $\langle \bar \psi \psi \rangle\to 0$, the chiral partner  becomes (approximately) degenerate:
\begin{eqnarray}
\mbox{$SU(2)_L\times SU(2)_R$ restoration limit}:
&&
\begin{cases}
\chi_\pi-\chi_\sigma\to0\\
\chi_{a_0}-\chi_{\eta}\to 0
\end{cases}
.
\end{eqnarray}
After the chiral restoration, the topological susceptibility is dominated by the $U(1)_A$ axial partner: 
$\chi_{\rm top}\sim\chi_{\eta}-\chi_{\sigma}$
($\chi_{\rm top}\sim\chi_{\pi}-\chi_{a_0}$),
so that $\chi_{\rm top}$ acts as the indicator for the breaking strength of $U(1)_A$ symmetry. It should be noted that
$\chi_{\rm top}$ trivially vanishes in the chiral limit ($m_l=0$) as seen from Eq.~(\ref{chitop_meson_sus}). In this limit, 
the topological susceptibility is no longer regarded as the indicator. This can also be understood by the fact that when $m_l=0$, the $\theta$ dependence of the generating functional in Eq.~(\ref{gene_func_mass_theta}) disappears, resulting in the vanishing topological susceptibility defined by a second derivative with respect to $\theta$.

When studying with a low-energy effective model, the analytical expression of~(\ref{chitop_meson_sus}) is useful for evaluating the topological susceptibility $\chi_{\rm top}$. Here, we show another expression of $\chi_{\rm top}$ so as to see contributions from the chiral condensate $\langle\bar{\psi}\psi\rangle$ clearly. That is, from the identity~(\ref{AWI_chiral}) one can rewrite Eq.~(\ref{chitop_meson_sus}) into
\begin{eqnarray}
\chi_{\rm top} = -\frac{m_l\langle\bar{\psi}\psi\rangle}{4}\delta_m\ .
\label{ChiTopQbarQ}
\end{eqnarray}
In this expression, the dimensionless quantity $\delta_m$ is defined by
\begin{eqnarray}
\delta_m \equiv 1-\frac{\chi_\eta}{\chi_\pi}\ , \label{DeltaM}
\end{eqnarray}
which measures the variation of the susceptibility functions $\chi_\pi$ and $\chi_\eta$. Equation~(\ref{ChiTopQbarQ}) indicates, indeed, that the topological susceptibility is proportional to the chiral condensate $\langle\bar{\psi}\psi\rangle$ and $\delta_m$; the explicit chiral-symmetry breaking is entangled with the $U(1)_A$ anomaly contribution captured by the quantity $\delta_m$ in $\chi_{\rm top}$. This structure plays an important role in determining the asymptotic behavior of $\chi_{\rm top}$ at sufficiently large $\mu_q$ where chiral symmetry is restored.

Furthermore, the Gell-Mann-Oakes-Renner (GOR) relation: $f_\pi^2 m_\pi^2=-m_l\langle\bar{\psi}\psi\rangle/2$~\cite{cheng1994gauge}, enables us to rewrite the topological susceptibility in Eq.~(\ref{ChiTopQbarQ}) as
\begin{eqnarray}
\chi_{\rm top}= \frac{f_\pi^2m_\pi^2}{2}
\delta_m\ ,
\label{chi_top_vev_wdelta}
\end{eqnarray}
where $f_\pi$ is the pion decay constant and $m_\pi$ is the pion mass. It is obvious from its derivation that Eq.~(\ref{chi_top_vev_wdelta}) holds model-independently.\footnote{The GOR relation $f_\pi^2 m_\pi^2=-m_l\langle\bar{\psi}\psi\rangle/2$ is derived model-independently but with an assumption that the two-point function of the pseudoscalar channel $D_\pi \delta^{ab} \equiv \int d^4x\langle 0|T (i\bar \psi\gamma_5 \tau^a_f \psi)(x) (i\bar \psi\gamma_5 \tau^b_f \psi)(0)|0\rangle{\rm e}^{-ip\cdot x}$ is dominated by the lightest pseudoscalar-meson pole
: $D_\pi \propto i/(p^2-m_\pi^2)$, as in the case of three-color QCD. Accordingly, the relation~(\ref{chi_top_vev_wdelta}) also holds upon the pole dominance of the lightest pseudoscalar meson.
} Notably, the quantity $\delta_m$ in the vacuum is solely determined by the masses of pion and $\eta$ meson as
\begin{eqnarray}
\delta_m \to 1-\frac{m_\pi^2}{m_\eta^2}\ , \label{mass_difference}
\end{eqnarray}
as long as we stick to the low-energy regime of QC$_2$D where $\chi_\pi\sim-i/m_\pi^2$ and $\chi_\eta\sim-i/m_\eta^2$ can apply. Therefore, Eq.~(\ref{chi_top_vev_wdelta}) implies that the topological susceptibility in the vacuum is expressed by three basic observables in low-energy QC$_2$D: $f_\pi$, $m_\pi$ and $m_\eta$. We note that $\delta_m\to1$ corresponds to the significantly large anomaly effects, while $\delta_m\to0$ implies no such effects.
We also note that 
Leutwyler and Smilga
obtained the following form based on 
the Chiral Perturbation Theory (ChPT) in three-color QCD~\cite{Leutwyler:1992yt}:
\begin{eqnarray}
\chi^{\rm (LS)}_{\rm top}= -\frac{m_l\langle\bar{\psi}\psi\rangle}{4}
=
\frac{f_\pi^2m_\pi^2}{2}\ ,
\label{LS_formula}
\end{eqnarray}
where the $\chi_\eta$ contributions are missing.
Indeed, in Eq.~(\ref{LS_formula}),
the large anomaly is accidentally taken into account: $\delta_m=1$.
Even in the case of QC$_2$D, the Leutwyler-Smilga relation was also found in ~\cite{Metlitski:2005db}. 




\section{
Low-energy effective-model description of two-flavor ${\rm QC_2D}$
}
\label{sec:LSM}

In ${\rm QC_2D}$, diquarks (antidiquarks) carrying the quark number $+2$ ($-2$) are treated as color-singlet baryons, namely, baryons become bosonic similarly to mesons. Accordingly, the so-called Pauli-Gursey $SU(4)$ symmetry, which enables us to treat both the baryons and mesons in a consistent way, emerges~\cite{Kogut:1999iv,Kogut:2000ek}. In this section, we briefly explain how the Pauli-Gursey $SU(4)$ symmetry manifests itself from QC$_2$D Lagrangian, and based on the symmetry we present the linear sigma model which describes couplings among the baryons and mesons.

Thanks to pseudoreal properties of the $SU(2)$ generators for color and Dirac spaces, $T_c^a=-\tau^2(T_c^a)^T\tau^2$ and $\sigma^i=-\sigma^2(\sigma^i)^T\sigma^2$ ($\sigma^i$ is the Pauli matrix in the Dirac space), one can show that the kinetc term of quarks coupling with gauge fields in QC$_2$D, i.e., the first piece in Eq.~(\ref{QCD_lag_theta}), is rewritten to 
\begin{eqnarray}
{\cal L}_{\rm Q_2CD}^{(\rm kin)}=
 \Psi^\dagger i\sigma^\mu D_\mu \Psi \ ,
 \label{QCD_lag_kin}
\end{eqnarray}
with $\sigma^\mu=({\bm 1}, \sigma^i)$, in the Weyl representation. In Eq.~(\ref{QCD_lag_kin}) the quark field $\Psi$ is given by a four-component column vector in the flavor space as
\begin{eqnarray}
\Psi=
\begin{pmatrix}
\psi_R\\
\tilde \psi_L
\end{pmatrix}
=
\begin{pmatrix}
u_R\\
d_R\\
\tilde u_L\\
\tilde d_L 
\end{pmatrix} \ ,
\end{eqnarray}
where $\psi_{L(R)}=\frac{1\mp\gamma_5}{2}\psi$ denotes the left-handed (right-handed) quark field and $\tilde \psi_{L(R)}$ is the conjugate one:
\begin{eqnarray}
\tilde \psi_{L(R)}=\sigma^2 \tau^2_c \psi_{L(R)}^*\ .
\end{eqnarray}
Equation~(\ref{QCD_lag_kin}) implies that the quark kinetic term in QC$_2$D is invariant under an $SU(4)$ transformation for the quark field as
\begin{eqnarray}
\Psi\to g \Psi\ ,
\end{eqnarray}
with $g \in SU(4)$. Thus, it is proven that $SU(2)_L\times SU(2)_R$ chiral symmetry in QC$_2$D is extended to the $SU(4)$ one which is often referred to as the Pauli-Gursey $SU(4)$ symmetry~\cite{Kogut:1999iv,Kogut:2000ek}.

Similarly to the kinetic part, the quark mass term, i.e., the second piece in Eq.~(\ref{QCD_lag_theta}), is also expressed in terms of the four-component field $\Psi$, which reads
\begin{eqnarray}
{\cal L}_{\rm QC_2D}^{(\rm mass)}= \frac{m_l}{2}\left( \Psi^T \sigma^2 \tau_c^2E \Psi
+ \Psi^\dagger \sigma^2 \tau_c^2E^T \Psi^*\right)\ .
 \label{QCD_lag_mass}
\end{eqnarray}
This term, however, breaks the Pauli-Gursey $SU(4)$ symmetry explicitly due to the presence of a symplectic matrix in the flavor space
\begin{eqnarray}
E=\left(
\begin{array}{cc}
0 & {\bm 1} \\
-{\bm 1} & 0 \\
\end{array}
\right) \ ,
\end{eqnarray}
in between the two quark fields. For this reason, 
the systematic treatment based on the viewpoint of the $SU(4)$ symmetry is spoiled by the quark masses.
To recover the systematics, we introduce a spurion field $\zeta_{\rm sp}$ which transforms as
\begin{eqnarray}
\zeta_{\rm sp}\to g\,\zeta_{\rm sp}\, g^T\ . \label{ChiTrans}
\end{eqnarray}
To construct the $SU(4)$-invariant Lagrangian, 
the quark mass term is promoted to the spurion term
\begin{eqnarray}
{\cal L}_{\rm QC_2D}^{\rm (sp)}
=
-\Psi^T\sigma^2\tau^2_c\zeta^\dagger_{\rm sp} \Psi
-\Psi^\dagger\sigma^2\tau^2_c\zeta_{\rm sp}\Psi^*\ .
\label{spterm_QCD}
\end{eqnarray}
In fact,
one can show that the quark mass term~(\ref{QCD_lag_mass}) is appropriately reproduced by taking the vacuum expectation value (VEV) of the spurion field as
\begin{eqnarray}
\langle \zeta_{\rm sp} \rangle =
\frac{m_l}{2} E\ . \label{ChiSP}
\end{eqnarray}

In what follows, we construct the linear sigma model to describe hadrons at the low-energy regime of QC$_2$D, based on the symmetries explained above. The fundamental building block of the linear sigma model in QC$_2$D is a $4\times4$ matrix field $\Sigma_{ij}$ whose symmetry properties are the same as those of a quark bilinear field $\Psi_j^T\sigma^2\tau_c^2\Psi_i$. That is, $\Sigma$ transforms as 
\begin{eqnarray}
\Sigma \to g \Sigma g^T\ ,
\end{eqnarray}
under the $SU(4)$ transformation. As explained in Ref.~\cite{Suenaga:2022uqn} in detail, the $\Sigma$ can be parameterized by low-lying hadrons in QC$_2$D as
\begin{eqnarray}
\Sigma=(S^a-iP^a)X^aE+(B^{\prime i}-iB^i)X^iE \ ,  \label{SigmaDef0}
\end{eqnarray}
where $S^a$, $P^a$, $B^i$ and $B^{\prime i}$ represent scalar mesons, pseudoscalar mesons, positive-parity diquark baryons and negative-parity diquark baryons, respectively. The $4\times4$ matrices $X^a$ and $X^i$ are generators of $U(4)$ defined by
\begin{eqnarray}
X^a&=&
\frac{1}{2\sqrt{2}}
\begin{pmatrix}
\tau^a_f&0\\
0& (\tau^a_f)^T
\end{pmatrix}
\;\;\;
(a=0,1,2,3)\ ,\nonumber\\
X^i&=&
\frac{1}{2\sqrt{2}}
\begin{pmatrix}
0&D_f^i\\
 (D_f^i)^\dagger&0
\end{pmatrix}
\;\;\;
(i=4,5)\ ,
\end{eqnarray}
where $\tau_f^0={\bm 1}_{2\times 2}$ in the flavor space, and
$D_f^i$ represent $D_f^4=\tau_f^2$ and $D_f^5=i\tau_f^2$. Following the parametrization given in Ref.~\cite{Suenaga:2022uqn}, we employ the following hadron assignment for $\Sigma$:
\begin{eqnarray}
\Sigma=
\frac{1}{2}
\begin{pmatrix}
0&-B^\prime+iB&\frac{\sigma-i\eta+a^0-i\pi^0}{\sqrt{2}}&a^+-i\pi^+\\
B^\prime-iB&0& a^--i\pi^-&
\frac{\sigma-i\eta-a^0+i\pi^0}{\sqrt{2}}\\
-\frac{\sigma-i\eta+a^0-i\pi^0}{\sqrt{2}}&
-a^-+i\pi^-&0&
-\bar B^\prime+i\bar B\\
-a^++i\pi^+&-\frac{\sigma-i\eta-a^0+i\pi^0}{\sqrt{2}}&\bar B^\prime-i\bar B&0
\end{pmatrix}\ ,  \label{SigmaDef}
\end{eqnarray}
where $\pi^0= P^3$ and $\pi^\pm=(P^1\mp iP^2)/\sqrt{2}$ are the pions, $\eta = P^0$ is the $\eta$ meson,
$\sigma=S^0$ is the iso-singlet scalar meson ($\sigma$ meson),
$a_0^0=S^3$ and $a_0^\pm=(S^1\mp i S^2)/\sqrt{2}$ are the iso-triplet scalar mesons ($a_0$ mesons),
$B=(B^5-iB^4)/\sqrt{2}$ [$\bar B=(B^5+iB^4)/\sqrt{2}$] is the positive-parity diquark baryon (the antidiquark baryon), and
$B^\prime=(B^{\prime5}-iB^{\prime4})/\sqrt{2}$ [$\bar B^\prime=(B^{\prime5}+iB^{\prime4})/\sqrt{2}$] is the negative-parity diquark baryon (the antidiquark baryon).

With the matrix $\Sigma$ given in Eq.~(\ref{SigmaDef}), our linear sigma model in QC$_2$D which respects the Pauli-Gursey $SU(4)$ symmetry is obtained as
\begin{eqnarray}
{\cal L}_{\rm LSM}
&=&
{\rm tr }[D_\mu\Sigma^\dagger D^\mu \Sigma ]
-V\ ,
\label{LSM_lag}
\end{eqnarray}
where the covariant derivative for $\Sigma$ is defined by
\begin{eqnarray}
D_\mu \Sigma= \partial_\mu \Sigma -i \mu_q \delta_{\mu0}(J\Sigma+ \Sigma J^T)
\;\;\;
\mbox{with}
\;\;\;
J=
\begin{pmatrix}
{\bm 1}&0\\
0&-{\bm1}
\end{pmatrix}.
\end{eqnarray}
Here, the quark chemical potential $\mu_q$ is incorporated 
in the covariant derivative through gauging the $U(1)_B$ baryon-number symmetry. 
Besides, $V$ represents potential terms describing interactions among the hadrons, which is separated into three parts as
\begin{eqnarray}
V=V_0+V_{\rm sp}+V_{\rm anom}\ . \label{VPot}
\end{eqnarray}
$V_0$ represents an invariant part under the Pauli-Gursey $SU(4)$ symmetry.
When we include contributions up to the fourth order of $\Sigma$ as widely done in the linear sigma model for three-color QCD, it takes the form of
\begin{eqnarray}
V_0=
m_0^2 {\rm tr}[\Sigma^\dagger \Sigma]+
\lambda_1({\rm tr}[\Sigma^\dagger \Sigma])^2
+\lambda_2{\rm tr}[(\Sigma^\dagger \Sigma)^2] \ ,
\end{eqnarray}
where $m_0^2$ is a mass parameter, and $\lambda_1$ and $\lambda_2$ are coupling constants controlling the strength of four point interactions. The second piece of Eq.~(\ref{VPot}), $V_{\rm sp}$, is the spurion term corresponding to ${\cal L}_{\rm QC_2D}^{\rm (sp)}$ in Eq.~(\ref{spterm_QCD}), which is given by
\begin{eqnarray}
V_{\rm sp}
=-\bar c\,{\rm tr}[ \zeta^\dagger_{\rm sp} \Sigma +\Sigma^\dagger \zeta_{\rm sp}]\ ,
\label{sp_LSM}
\end{eqnarray}
where the parameter $\bar c$ is real, and has the mass dimension two. Although the $V_{\rm sp}$ is invariant under the $SU(4)$ transformation thanks to Eq.~(\ref{ChiTrans}), the spurion field $\chi_{\rm sp}$ must be replaced by its VEV in Eq.~(\ref{ChiSP}) 
so as to incorporate the effect of the finite quark mass
in a final step for evaluating physical observables.

The last piece in the potential~(\ref{VPot}), $V_{\rm anom.}$, includes $U(1)_A$ anomalous contributions which is responsible for the gluonic part in the non-conservation law of the axial current: the second term of the right-hand side (RHS) of Eq.~(\ref{AxialD}). Within our present model, the $U(1)_A$ anomalous term is expressed by the Kobayashi-Maskawa-'t Hooft (KMT)-type interaction,~\cite{Kobayashi:1970ji,Kobayashi:1971qz,tHooft:1976snw,tHooft:1976rip}
\begin{eqnarray}
V_{\rm anom}
=-c({\rm det}\Sigma +\det\Sigma^\dagger)\ .
\label{KMT}
\end{eqnarray}
As demonstrated below, this anomalous term generates a mass difference between the pion and $\eta$ meson in the vacuum~\cite{Suenaga:2022uqn}, and plays an important role in driving a finite topological susceptibility. It should be noted that the KMT-type interaction 
is described by four-incoming and four-outgoing quarks owing to  
the quark bilinear field $\Sigma_{ij}\sim\bar \Psi^T_j \sigma^2\tau_c^2\Psi_i$ based on the Pauli-Gursey $SU(4)$ symmetry. Thus, in hadronic-level diagrams,
$V_{\rm anom}$ represents four-point interactions.

\section{
Topological susceptibility at low energy 
}
\label{sec:Analysis}

General expressions and characteristics
of the topological susceptibility in QC$_2$D have been reviewed in Sec.~\ref{sec:chitop_derivation}, and the linear sigma model, which describes hadrons in the low-energy regime of QC$_2$D, has been invented in Sec.~\ref{sec:LSM}. In this section, we explain our strategy to evaluate the topological susceptibility within our linear sigma model through matching with the underlying QC$_2$D theory.

\subsection{
Matching between low-energy effective model and underlying QC$_2$D
}
\label{sec:Matching}

In Sec.~\ref{sec:LSM} we have constructed the linear sigma model in order to describe the hadrons as low-energy excitations of underlying QC$_2$D. 
On the basis of the concept of the low-energy effective theory,  the linear sigma model is equivalent to
 QC$_2$D in the low-energy regime
 through the generating functional:
\begin{eqnarray}
Z_{\rm QC_2D} = Z_{\rm LSM}
=
\int [d\Sigma]
\exp\left(
i\int d^4x
{\cal L}_{\rm LSM }
\right)\ .
\label{matching_gene}
\end{eqnarray}
In this subsection, we discuss the matching of the physical quantities between the linear sigma model and underlying QC$_2$D based on Eq.~(\ref{matching_gene}). Note that
we neglect spin-$1$ hadronic excitations such as the $\rho$ meson in the low-energy theory $Z_{\rm LSM}$, even though 
the mass spectrums of spin-$1$ mesons coexist with that of spin-$0$ mesons in the low-energy regime~\cite{Murakami:2022lmq}. 
This is because 
the topological susceptibility is evaluated by only the susceptibility functions $\chi_\pi$ and $\chi_\eta$ as in Eq.~(\ref{chitop_meson_sus}), which do not include spin-$1$ operators. 
The spin-$1$ hadronic excitations would hardly contribute to the following results. 
 

In a similar way to Eq.~(\ref{VQC2D}), 
the effective action of the linear sigma model is given by
\begin{eqnarray}
\Gamma_{{\rm LSM}}= -i \ln Z_{{\rm LSM} }\ .
\end{eqnarray}
From the equivalence in Eq.~(\ref{matching_gene}), 
we have the following matching condition in terms of $\Gamma$'s:
\begin{eqnarray}
\Gamma_{{\rm QC_2D}} = \Gamma_{{\rm LSM}}\ ,
\label{matching_vacu}
\end{eqnarray}
Here, we emphasize that both the effective actions $\Gamma_{{\rm QC_2D}}$ and $\Gamma_{{\rm LSM}}$ depend on the spurion field $\zeta_{\rm sp}$ commonly to maintain the systematics of $SU(4)$ symmetry. In general, $\zeta_{\rm sp}$ takes the form of
\begin{eqnarray}
\zeta_{\rm sp}=
( \zeta_{ S}^a-i \zeta_{ P}^a)X^aE+(
\zeta_{{B}^{\prime }}^i-i \zeta_{B}^i)X^iE\ , \label{Spurion}
\end{eqnarray}
where
$\zeta_{S}^a$ ($\zeta_{P}^a$) are scalar (pseudoscalar) source fields, and $\zeta_{B}^i$ ($\zeta_{B^{\prime }}^i$) are source fields associated with the positive-parity (negative-parity) diquark baryons.

Taking functional derivatives with respect to 
the source fields in both sides of Eq.~(\ref{matching_vacu}), the matching between the linear sigma model and underlying QC$_2$D can be done. For instance, functional derivatives with respect to the scalar source field $ \zeta_S^{0}$ yield\footnote{The quark mass term in Eq.~(\ref{QCD_lag_theta}) is now replaced by the spurion term~(\ref{spterm_QCD}).}
\begin{eqnarray}
\langle \bar \psi \psi \rangle
&=&
-\sqrt{2}
\frac{\delta \Gamma_{{\rm QC_2D}}}{\delta \zeta_S^{0}(x) }
\Biggl|_{\zeta_{\rm sp}=\langle\zeta_{\rm sp}\rangle}
\nonumber\\
&=&
-\sqrt{2}
\frac{\delta \Gamma_{{\rm LSM}}}{\delta \zeta_S^{0}(x) }
\Biggl|_{\zeta_{\rm sp}=\langle\zeta_{\rm sp}\rangle}
=-\sqrt{2}\bar c \langle \sigma\rangle \ .
\label{matched_qcon}
\end{eqnarray}
This implies that the chiral condensate $\langle\bar{\psi}\psi\rangle$ serving as an order parameter of the spontaneous breakdown of chiral symmetry is evaluated by a VEV of $\sigma$ meson within the linear sigma model. Moreover, one can see that the chiral condensate is rewritten as
\begin{eqnarray}
\langle \bar \psi \psi \rangle
=
\left\langle
\left(-\frac{1}{2}
\Psi^T\sigma^2\tau_c^2E\Psi
+{\rm h.c.}
\right)\right\rangle\ .
\end{eqnarray}
This shows that the chiral condensate is invariant under a transformation with $h$ which is an element of $Sp(4)$ belonging to a subgroup of $SU(4)$,
\begin{eqnarray}
h^TEh = E \ .
\end{eqnarray}
Hence, the symmetry-breaking pattern caused by the chiral condensate is $SU(4)\to Sp(4)$.

Likewise, when we take functional derivatives of Eq.~(\ref{matching_vacu}) with respect to $\zeta_B^5$, the following equivalence is obtained:
\begin{eqnarray}
\left\langle\left(
-\frac{i}{2}\psi^TC\gamma_5\tau_c^2\tau_f^2\psi
+{\rm h.c.}
\right)\right\rangle
&=&
-\sqrt{2}
\frac{\delta \Gamma_{{\rm QC_2D}}}{\delta \zeta_B^{5}(x) }
\Biggl|_{\zeta_{\rm sp}=\langle\zeta_{\rm sp}\rangle}
\nonumber\\
&=&
-\sqrt{2}
\frac{\delta \Gamma_{{\rm LSM}}}{\delta \zeta_B^{5}(x) }
\Biggl|_{\zeta_{\rm sp}=\langle\zeta_{\rm sp}\rangle}
=
-\sqrt{2}
\bar c\langle B^5 \rangle\ , \label{MatchDelta}
\end{eqnarray}
with the charge-conjugation operator $C=i\gamma^2\gamma^0$. This equation indicates that the diquark condensate $\langle\psi^T C\gamma_5\tau_c^2\tau_f^2\psi\rangle$, which plays a role of the order parameter for the emergence of the baryon superfluid phases, is mimicked by a VEV of the diquark baryon field $B^5$ in the linear sigma model. 
Since $B^5$ carries a finite quark number, the quark-number conservation no longer holds in the superfluid phase. It should be noted that 
the common coefficient $\bar{c}$ in Eqs.~(\ref{matched_qcon}) and~(\ref{MatchDelta}) is the result of the Pauli-Gursey $SU(4)$ symmetry
 which combines mesons and diquark baryons into the single multiplet.

At zero temperature, low-energy effective theories such as the linear sigma model undergo 
the baryon superfluid phase transition 
at 
the half value of the vacuum pion mass: $\mu^{\rm cr}_q=m^{\rm vac}_\pi/2$~\cite{Kogut:2000ek,Ratti:2004ra,Suenaga:2022uqn}. Below this critical chemical potential, only the hadronic phase, where no diquark condensates emerge, is realized, and all thermodynamic quantities do not change against increment of $\mu_q$. This stable behavior is often referred to as the Silver-Braze property, and lattice simulations also support it~\cite{Iida:2022hyy}. 
Above the critical chemical potential $\mu_q^{\rm cr}$, 
the baryon superfluid phase transition occurs
and accordingly, the baryonic density also arises there. 
Meanwhile, in the baryon superfluid phase, the chiral condensate begins to decrease with increasing the baryonic density,
resulting in the (partial) restoration of chiral symmetry~\cite{Kogut:2000ek,Ratti:2004ra,Suenaga:2022uqn}.

In what follows, we use 
\begin{eqnarray}
\sigma_0
\equiv
\langle \sigma \rangle\ , \ \ \Delta\equiv \langle B^5 \rangle\ ,
\label{delcon}
\end{eqnarray}
to refer to the VEVs, where the phase of $\langle B^5 \rangle$
has been chosen to make $\Delta$ real.

In Eqs.~(\ref{matched_qcon}) and~(\ref{MatchDelta}), we have demonstrated 
how the QCD observables for VEVs of single local operators
are matched with physical quantities of the linear sigma model: the chiral condensate and diquark condensate.
The matching can be also done for two-point correlation functions by taking second functional derivatives in Eq.~(\ref{matching_vacu}) with respect to the source fields.
In fact, by performing functional derivatives appropriately, one can find that the $\eta$-meson and pion susceptibility functions, $\chi_\eta$ and $\chi_\pi$ defined in Eqs.~(\ref{eta_sus_QCD}) and~(\ref{PiSusQCD}), are related to two-point functions of the $\eta$ meson and pion in the linear sigma model, respectively, as
\begin{eqnarray}
\chi_\eta&=&-2i\int d^4x 
\frac{\delta^2 \Gamma_{{\rm QC_2D}}}{\delta \zeta_P^{0}(x) \delta \zeta_P^{0}(0) }
\Biggl|_{\zeta_{\rm sp}=\langle\zeta_{\rm sp}\rangle}
\nonumber\\
&=&-2i\int d^4x 
\frac{\delta^2 \Gamma_{{\rm LSM}}}{\delta \zeta_P^{0}(x) \delta \zeta_P^{0}(0) }
\Biggl|_{\zeta_{\rm sp}=\langle\zeta_{\rm sp}\rangle}
=2\bar c^2\int d^4x\langle 0|T\eta (x) \eta(0) |0 \rangle\ ,
\label{eta_sus_LSM}
\end{eqnarray}
and
\begin{eqnarray}
\chi_\pi\delta^{ab}&=&
-2i\int d^4x 
\frac{\delta^2 \Gamma_{{\rm QC_2D}}}{\delta \zeta_P^{a}(x) \delta \zeta_P^{b}(0) }
\Biggl|_{\zeta_{\rm sp}=\langle\zeta_{\rm sp}\rangle}
\nonumber\\
&=&-2i\int d^4x 
\frac{\delta^2 \Gamma_{{\rm LSM}}}{\delta \zeta_P^{a}(x) \delta \zeta_P^{b}(0) }
\Biggl|_{\zeta_{\rm sp}=\langle\zeta_{\rm sp}\rangle}
=2\bar c^2\int d^4x\langle 0|T\pi^a (x) \pi^b(0) |0 \rangle\;\;\;
(\mbox{for }a,b=1,2,3)\ . \nonumber\\
\label{pi_sus_LSM}
\end{eqnarray}
Using these matching equations, we will present analytic expressions of the meson susceptibility functions within our linear sigma model.



\subsection{
Topological susceptibility across baryon superfluid phase transition
}
\label{sec:TSinLSM}

In this subsection, we proceed with analytic evaluation of the topological susceptibility from the linear sigma model.


In this work, we employ the mean-field approximation where loop corrections of hadronic fluctuations have not been taken into account. The effective potential of the linear sigma model at the tree level is evaluated as
\begin{eqnarray}
V_{\rm mean}=
-2\mu_q^2\Delta^2+\frac{m_0^2}{2}(\sigma_0^2+\Delta^2)
+\frac{8\lambda_1+2\lambda_2-c}{32}(\sigma_0^2+\Delta^2)^2
-\sqrt{2}m_l\bar c\sigma_0\ .
\label{effpot_LSM}
\end{eqnarray}
In this potential the VEV of the spurion field~(\ref{ChiSP}) as well as the mean fields~(\ref{delcon}) are inserted. In Eq.~(\ref{effpot_LSM})
the quark chemical potential $\mu_q$ appears in a quadratic term of $\Delta$ with a negative sign, indicating that the larger value of $\mu_q$ yields nonzero $\Delta$ leading to the baryon superfluid phase as mentioned in Sec.~\ref{sec:Matching}. 
The vacuum configurations are determined by stationary conditinos of $V_{\rm mean}$ with respect to $\sigma_0$ and $\Delta$: 
\begin{eqnarray}
\frac{\partial V_{\rm mean}}{\partial \sigma_0} = 0 \ , \ \ \frac{\partial V_{\rm mean}}{\partial \Delta} = 0\ ,
\label{st_condiitons}
\end{eqnarray}
and hadrons appear as fluctuation modes upon the vacuum characterized by the conditions~(\ref{st_condiitons}). In this description, hadron masses are evaluated by quadratic terms of the fluctuations in the Lagrangian~(\ref{LSM_lag}) with $\sigma_0$ and $\Delta$ included. For instance, the pion mass reads
\begin{eqnarray}
m_\pi^2 = m_0^2+\frac{8\lambda_1+2\lambda_2-c}{8} (\sigma_0^2+\Delta^2) = \frac{\sqrt{2}m_l \bar c}{\sigma_0}\ .
\label{pion_mass}
\end{eqnarray}
 We note that the second equality in Eq.~(\ref{pion_mass}) is obtained by considering
 the stationary condition of $\sigma_0$ in Eq.~(\ref{st_condiitons}).

When we approximate the pion two-point function $\langle0|T\pi^a(x)\pi^b(0)|0\rangle$ at the tree level in the linear sigma model,
the pion susceptibility function in Eq.~(\ref{pi_sus_LSM}) is evaluated to be 
\begin{eqnarray}
\chi_\pi=
-2i\bar c^2 \frac{1}{m_{\pi}^2}\ .
\label{matched_chipi}
\end{eqnarray}
Similarly, we employ the tree-level approximation for the $\eta$-meson two-point function $\langle0|T\eta(x)\eta(0)|0\rangle$. However, since 
the violation of $U(1)_B$ baryon-number symmetry in the baryon superfluid phase causes the mixing among $\eta$ meson, the negative-parity diquark $B'$ and antidiquark $\bar{B}'$ (or equivalently $\eta$, $B'_4$ and $B'_5$),
the two-point function of the $\eta$ meson is not simply given by $-i/m_\eta^2$ where the $\eta$ mass is read from  $\eta^2$ term of the $\eta$ fluctuation from the vacuum.
By taking into account the mixing structure, 
the inverse propagator matrix for the $\eta$ - $B'_4$ - $B'_5$ sector in the momentum space at the rest frame ${\bm p}={\bm 0}$ is obtained as
\begin{eqnarray}
i{\bm D}^{-1} = i\left(
\begin{array}{ccc}
D_{\eta\eta} & D_{\eta B_4'} & D_{\eta B_5'} \\
D_{B_4'\eta} & D_{B_4'B_4'} & D_{B_4'B_5'} \\
D_{B_5'\eta} & D_{B_5'B_4'} & D_{B_5'B_5'} \\
\end{array}
\right)^{-1} = \left(
\begin{array}{ccc}
p_0^2-m_\eta^2 & 0 & -m_{B_5'\eta}^2 \\
0 & p_0^2-m_{B_4'}^2 &4i\mu_q p_0 \\
-m_{B_5'\eta}^2 & -4i\mu_qp_0 & p_0^2-m_{B_5'}^2 \\
\end{array}
\right) \ , \label{DInvMatrix}
\end{eqnarray}
where we have defined the two-point functions $D_{XY}$ by $D_{XY} = {\rm F.T.}\langle 0|TX(x)Y(0)|0\rangle$ with $X,Y=\eta,B_4'$ and $B_5'$, and the mass parameters read
\begin{eqnarray}
m_\eta^2 &=& m_\pi^2+\frac{\lambda_2}{2}\Delta^2+\frac{c}{4}(2\sigma_0^2+\Delta^2)\ , \nonumber\\
m_{B_5^\prime}^2&=&m_\pi^2-4\mu_q^2
+\frac{\lambda_2}{2}\sigma_0^2+\frac{c}{4}(\sigma_0^2+2\Delta^2)\ ,\nonumber\\
m^2_{B_5^\prime \eta}&=&\frac{2\lambda_2 -c}{4}\sigma_0\Delta\ . \label{MEtaB5}
\end{eqnarray}
Thus, inverting the matrix~(\ref{DInvMatrix}), one can find that $D_{\eta\eta}$ which is of interest now takes the form of
\begin{eqnarray}
D_{\eta\eta}(p_0) = \sum_{i=1,2,3}\frac{iZ_{\phi_i}}{p_0^2-m_{\phi_i}^2}\ .\label{DEtaMix}
\end{eqnarray}
In this expression, $m_{\phi_1}$, $m_{\phi_2}$ and $m_{\phi_3}$ represent mass eigenvalues of the $\eta$ - $B'_4$ - $B'_5$ sector, where the subscripts $\phi_1$, $\phi_2$ and $\phi_3$ stand for the corresponding eigenstates with which the masses satisfy $m_{\phi_1}>m_{\phi_2}>m_{\phi_3}$. The renormalization constants $Z_{\phi_i}$ in Eq.~(\ref{DEtaMix}) are evaluated by
\begin{eqnarray}
Z_{\phi_1} &=& \frac{{\cal N}_{\eta\eta}(m_{\phi_1})}{(m_{\phi_1}^2-m_{\phi_2}^2)(m_{\phi_1}^2-m_{\phi_3}^2)}\ , \nonumber\\
Z_{\phi_2} &=& \frac{{\cal N}_{\eta\eta}(m_{\phi_2})}{(m_{\phi_2}^2-m_{\phi_1}^2)(m_{\phi_2}^2-m_{\phi_3}^2)}\ , \nonumber\\
Z_{\phi_3} &=& \frac{{\cal N}_{\eta\eta}(m_{\phi_3})}{(m_{\phi_3}^2-m_{\phi_1}^2)(m_{\phi_3}^2-m_{\phi_2}^2)}\ , 
\end{eqnarray}
with
\begin{eqnarray}
{\cal N}_{\eta\eta}(p_0) = p_0^4-16\mu_q^2p_0^2-(m_{B_4'}^2+m_{B_5'}^2)p_0^2+m_{B_4'}^2m_{B_5'}^2\ .
\end{eqnarray}
We note that the constants satisfy a condition $Z_{\phi_1} + Z_{\phi_2} + Z_{\phi_3}=1$ reflecting the fraction conservation. 
Therefore, 
$Z_{\phi_i}$ correspond to the proportion of the mass eigenstates $\phi_i$
in the two-point function $D_{\eta\eta}$ 
while the information on the respective pole positions is 
read from
$1/(p_0^2-m_{\phi_i}^2)$ in Eq.~(\ref{DEtaMix}).
\footnote{In Eq.~(\ref{DEtaMix}) we have expressed $D_{\eta\eta}(p_0)$ in terms of three contributions of $\phi_1$, $\phi_2$ and $\phi_3$ so as to see roles of the mass eigenstates clearly. In the low-energy limit $p_0=0$, $D_{\eta\eta}(0)$ is of course equivalent to the simple form of
\begin{eqnarray}
D_{\eta\eta}(0) = -i\frac{m_{B_5'}^2}{m_\eta^2 m_{B_5'}^2-m_{B_5'\eta}^4}\ ,
\end{eqnarray}
which can be straightforwardly derived by evaluating the inverse matrix of $\eta$ - $B'_5$ sector of Eq~(\ref{DInvMatrix}).}

By using the $\eta$-meson propagator based on the mass eigenstates $\phi_i$ in Eq.~(\ref{DEtaMix}),
the $\eta$-meson susceptibility function is evaluated as
\begin{eqnarray}
\chi_\eta = 2\bar{c}^2D_{\eta\eta}(0) =\sum_{i=1,2,3}\chi_{\phi_i}\ , \label{ChiEtaSum}
\end{eqnarray}
with
\begin{eqnarray}
\chi_{\phi_i} \equiv-2i\bar{c}^2\frac{Z_{\phi_i}}{m_{\phi_i}^2}\ . \label{ChiEta}
\end{eqnarray}
Since the susceptibility is defined at the low-energy limit: $p_0=0$,
the susceptibilities $\chi_{\phi_i}$ in Eq.~(\ref{ChiEta}) are written by $Z^2_{\phi_i}/m^2_{\phi_i}$ with the constant $-2i\bar{c}^2$. Therefore, the strength of $\chi_\eta$
is controlled by the combination of the renormalization constants $Z_{\phi_i}$ and
the mass eigenvalues $m_{\phi_i}$.

\section{
Fate of topological susceptibility in dense QC$_2$D
}
\label{sec:Numerical}

With the help of the susceptibility functions $\chi_\pi$ and $\chi_\eta$ obtained in Eqs.~(\ref{matched_chipi}) and~(\ref{ChiEtaSum}), the topological susceptibility $\chi_{\rm top}$ is evaluated within our linear sigma model from Eq.~(\ref{chitop_meson_sus}):
\begin{eqnarray}
\chi_{\rm top} = \frac{im_l^2}{4}
\left(
\chi_\pi - \sum_{i=1,2,3}\chi_{\phi_i}\right)
\ .
\end{eqnarray}
In this section, based on it, we show the numerical results of $\chi_{\rm top}$ at finite $\mu_q$.


\subsection{
$U(1)_A$ anomaly contribution and $\mu_q$ dependence of topological susceptibility
}
\label{sec:NumericalSusceptibility}

As explained at the end of Sec.~\ref{sec:chitop_derivation}, the topological susceptibility is substantially controlled by the $U(1)_A$ axial anomaly, i.e., the mass difference between the $\eta$ meson and pion in low-energy QC$_2$D. For this reason, we particularly investigate $\chi_{\rm top}$ at finite $\mu_q$ with the two cases of $m^{\rm vac}_\eta/m^{\rm vac}_\pi=1$ and $m^{\rm vac}_\eta/m^{\rm vac}_\pi=1.5$. The former corresponds to vanishing anomaly effects for the hadron spectrum, while the latter implies the substantial anomaly effects.

\begin{table}[htbp]
\begin{center}
  \begin{tabular}{c|ccccc} \hline
Mass ratio & $c$ & $\lambda_1$ & $\lambda_2$ & $m_0^2$ & $m_l\bar{c}/2$ \\ \hline 
$m^{\rm vac}_\eta/m^{\rm vac}_\pi=1$ & $0$ & $0$ & $65.6$ & $-(693\, {\rm MeV})^2$ & $(364\, {\rm MeV})^3$ \\
$m^{\rm vac}_\eta/m^{\rm vac}_\pi=1.5$ & $21.8$ & $0$ & $54.7$ & $-(373\, {\rm MeV})^2$ & $(364\, {\rm MeV})^3$ \\ \hline
 \end{tabular}
\caption{Fixed parameters for $m^{\rm vac}_\eta/m^{\rm vac}_\pi=1$ and $m^{\rm vac}_\eta/m^{\rm vac}_\pi=1.5$. }
\label{tab:Parameters}
\end{center}
\end{table}

When we fix the mass ratio $m^{\rm vac}_\eta/m^{\rm vac}_\pi$, there remain four parameters to be determined. As inputs, we employ $m_\pi^{\rm vac}=738$ MeV and $m_{B'(\bar{B}')}^{\rm vac}=1611$ MeV from the recent lattice data~\cite{Murakami:2022lmq}. Besides, based on the previous work~\cite{Suenaga:2022uqn}, $\sigma^{\rm vac}_0=250$ MeV is used as another input as a typical value. 
For the last constraint, we take $\lambda_1=0$ corresponding to the large $N_c$ limit since $\lambda_1$ term includes a double trace in the flavor space. With those inputs, the model parameters are fixed as in Table.~\ref{tab:Parameters}. The table indicates that $m^{\rm vac}_\eta$ becomes larger than $m^{\rm vac}_\pi$ only when $c\neq0$. In other words, the KMT-type interaction mimicking the $U(1)_A$ anomaly effects in the linear sigma model generates the mass difference between the $\eta$ meson and pion, as expected from underlying QC$_2$D.

\begin{figure}[H] 
\begin{tabular}{cc}
\begin{minipage}{0.5\hsize}
\begin{center}
    \includegraphics[width=7.2cm]{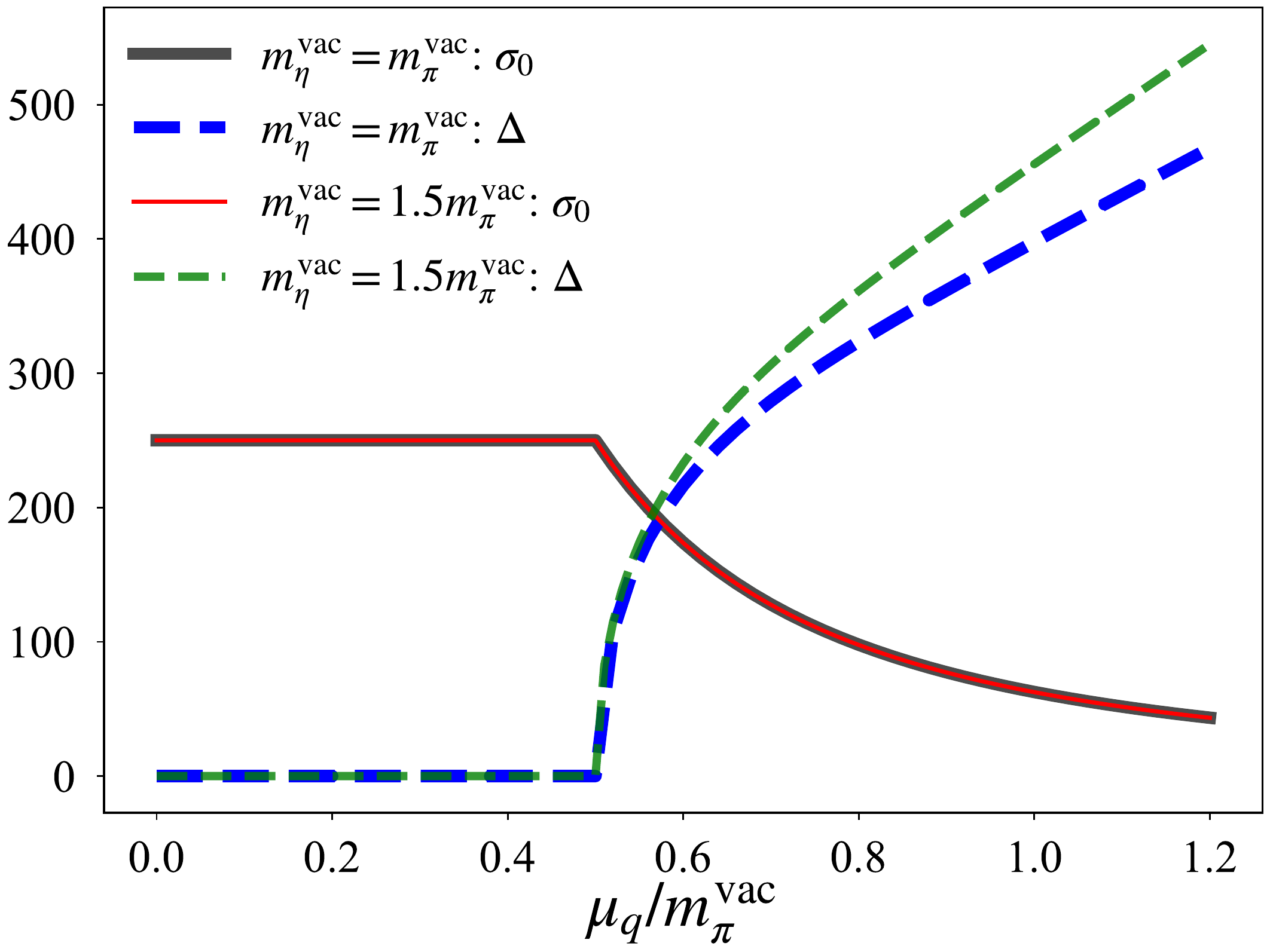}
    \subfigure{(a)}
\end{center}
\end{minipage}
\begin{minipage}{0.5\hsize}
\begin{center}
    \includegraphics[width=7.2cm]{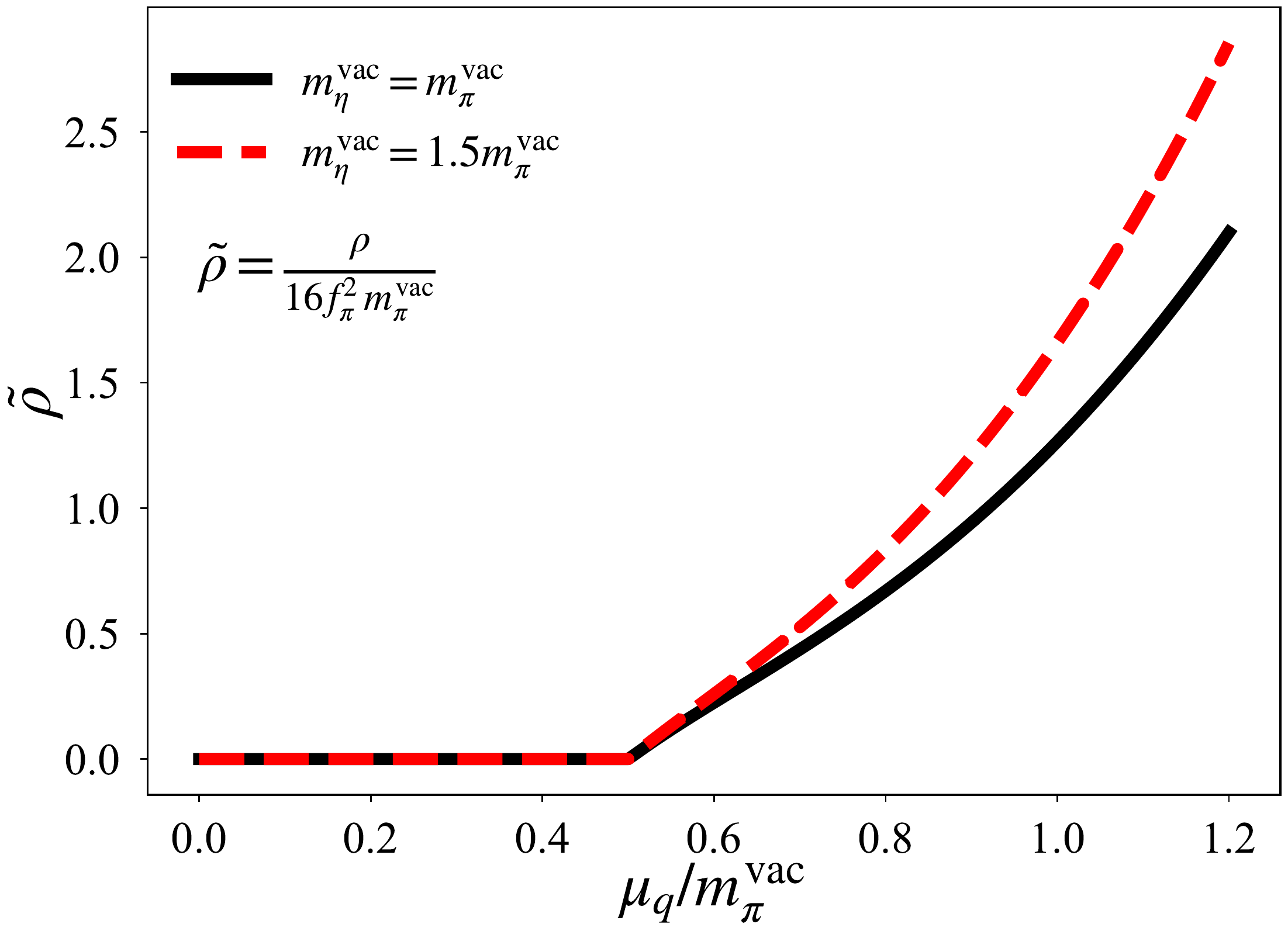}
    \subfigure{(b)}
\end{center}
\end{minipage}
\end{tabular}
\caption{Chemical potential
$\mu_q$ dependences of chiral condensate $\sigma_0$ and diquark condensate $\Delta$ (a)
and that of the baryon density $\rho$ normalized by $16f_\pi^2m_\pi^{\rm vac}$ (b).
}
\label{expectations}
\end{figure}

In order to demonstrate typical phase structures at zero temperature and finite chemical potential described by the present linear sigma model, we depict $\mu_q$ dependences of 
the chiral condensate $\sigma_0$ and diquark condensate $\Delta$ in the panel (a) of Fig.~\ref{expectations} for the two parameter sets of Table~\ref{tab:Parameters}. This figure clearly shows that the baryon superfluid phase emerges from $\mu_q^{\rm cr}=m_\pi^{\rm vac}/2$, and accordingly chiral symmetry begins to be restored. The mean field $\sigma_0$ decreases in the superfluid phase independently of the strength of the $U(1)_A$ anomaly effects, whereas the anomaly accelerates the increment of $\Delta$ there. We note that the smooth reduction of $\sigma_0$ in the superfluid phase is analytically evaluated as
\begin{eqnarray}
\sigma_0=\frac{m_l\bar{c}}{2\sqrt{2}}\mu_q^{-2}\ , \label{SigmaReduction}
\end{eqnarray}
from the stationary conditions in Eq.~(\ref{st_condiitons}). 
We also note that in the case of the nonlinear representation of the Nambu-Goldstone (NG) bosons, the vacuum manifold of the $SU(4)$ symmetry breaking is constrained as ``$\sigma_0^2+\Delta^2=({\rm constant})$'' at the tree level which was found in Refs.~\cite{Kogut:1999iv,Kogut:2000ek}. However, this is not the case in the linear representation~\cite{Suenaga:2022uqn}.

Incidentally,
Fig.~\ref{expectations} also depicts the $\mu_q$ dependence of the baryon number density ($\rho=4\Delta^2 \mu_q$) normalized by $16f_\pi^2m_\pi^{\rm vac}$ in panel (b).
The baryon number density is generated after reaching the baryon superfluid phase. 
Owing to the increment of $\Delta$, the baryon number density is enhanced by the $U(1)_A$ anomalous contribution.

\begin{figure}[H] 
\begin{tabular}{cc}
\begin{minipage}{0.5\hsize}
\begin{center}
    \includegraphics[width=7.2cm]{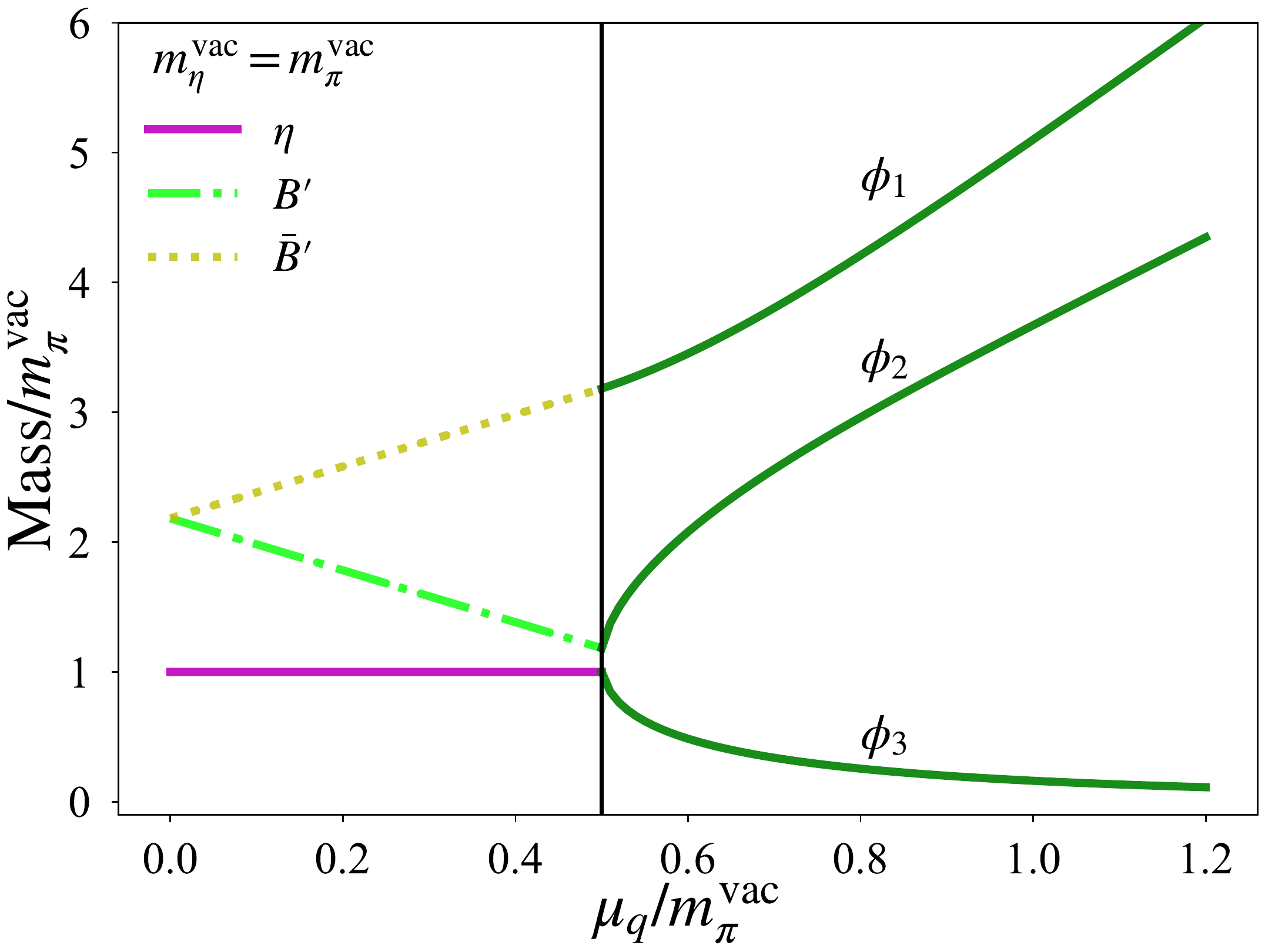}
    \subfigure{(a)}
\end{center}
\end{minipage}
\begin{minipage}{0.5\hsize}
\begin{center}
    \includegraphics[width=7.2cm]{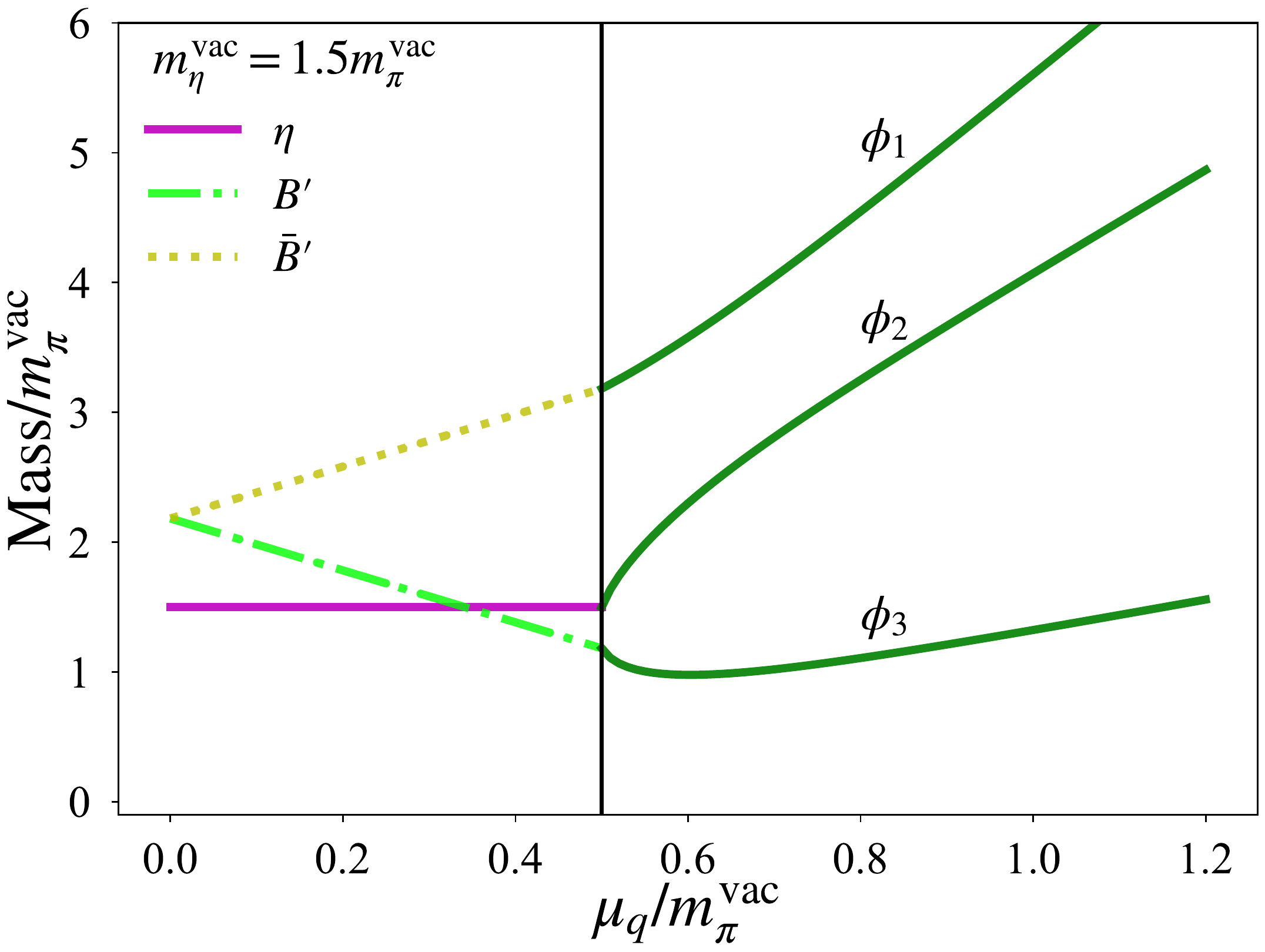}
    \subfigure{(b)}
\end{center}
\end{minipage}
\end{tabular}
\caption{
$\mu_q$ dependence of the masses of $\eta$ - $B'$ - $\bar{B}'$ sector for $m_\eta^{\rm vac}/m_\pi^{\rm vac}=1$ (a) and $m_\eta^{\rm vac}/m_\pi^{\rm vac}=1.5$ (b). In the baryon superfluid phase, $m_{\phi_1}$, $m_{\phi_2}$ and $m_{\phi_3}$ correspond to the green curves from above. In this figure the hadron masses are scaled by $m_\pi^{\rm vac}$. 
}
\label{fig:Mass}
\end{figure}

\begin{figure}[H] 
\begin{tabular}{cc}
\begin{minipage}{0.5\hsize}
\begin{center}
    \includegraphics[width=7.2cm]{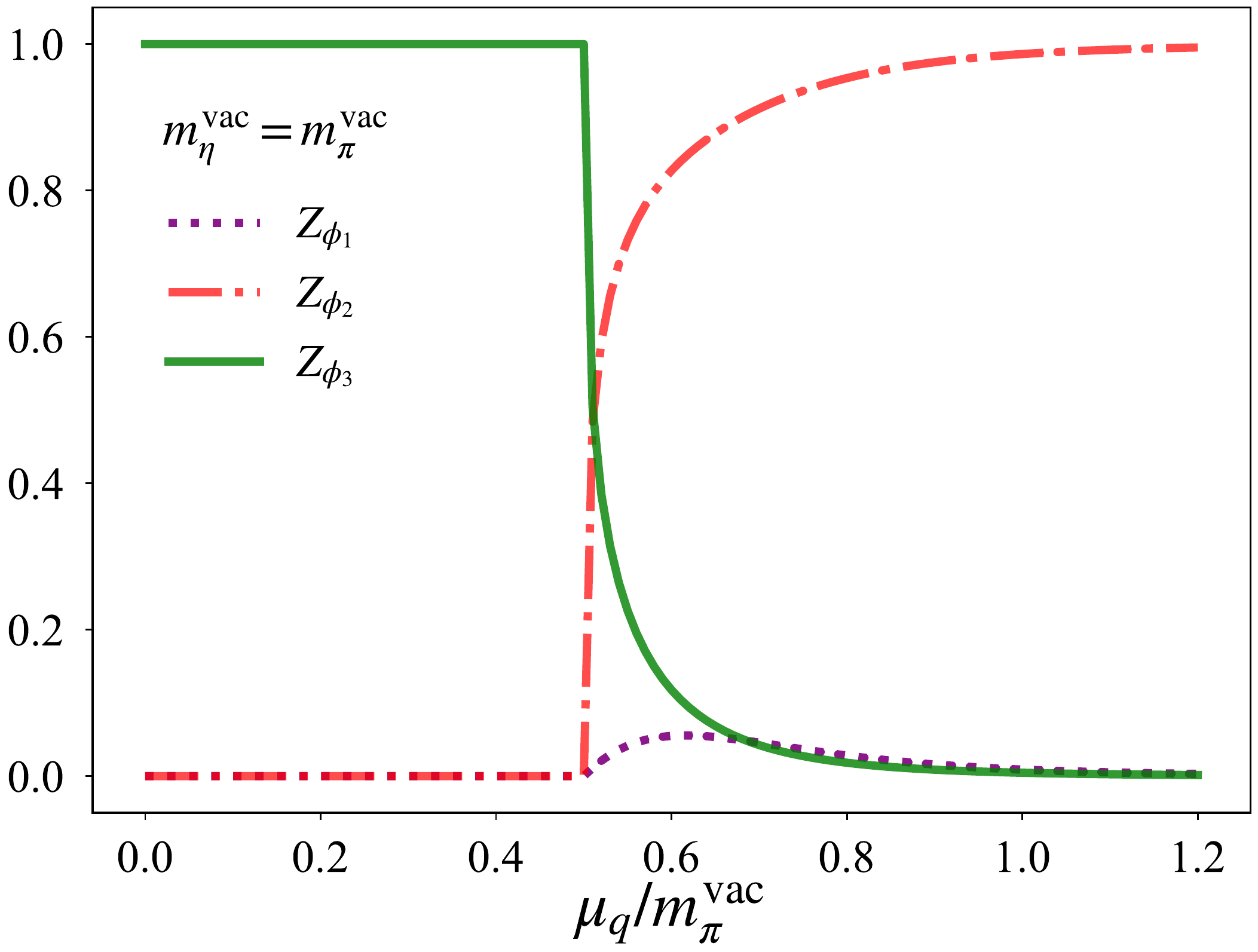}
    \subfigure{(a)}
\end{center}
\end{minipage}
\begin{minipage}{0.5\hsize}
\begin{center}
    \includegraphics[width=7.2cm]{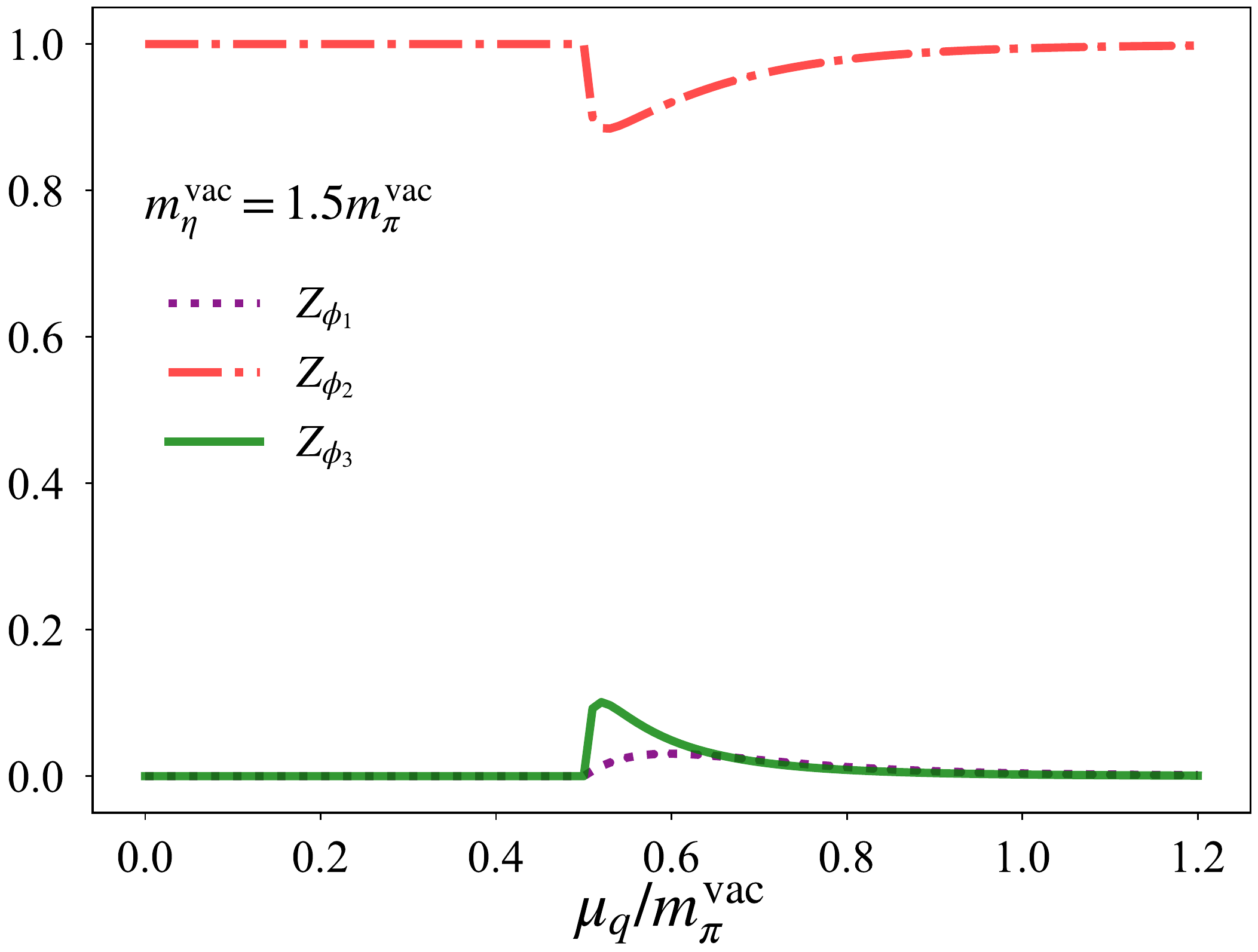}
    \subfigure{(b)}
\end{center}
\end{minipage}
\end{tabular}
\caption{
$\mu_q$ dependence of the renormalization constants $Z_{\phi_i}$ for $m_\eta^{\rm vac}/m_\pi^{\rm vac}=1$ (a) and $m_\eta^{\rm vac}/m_\pi^{\rm vac}=1.5$ (b). }
\label{fig:Ratio}
\end{figure}

For later convenience, here we comment on the masses $m_{\phi_i}$ and renormalization constants $Z_{\phi_i}$ across the phase transition. Displayed in Fig.~\ref{fig:Mass} is $\mu_q$ dependences of the masses of $\eta$ - $B'$ - $\bar{B}'$ sector for $m_\eta^{\rm vac}/m_\pi^{\rm vac}=1$ (a) and $m_\eta^{\rm vac}/m_\pi^{\rm vac}=1.5$ (b). In the baryon superfluid phase, $m_{\phi_1}$, $m_{\phi_2}$ and $m_{\phi_3}$ correspond to the green curves from above: $m_{\phi_i}$ are ordered from the largest mass,   $m_{\phi_1}>m_{\phi_2}>m_{\phi_3}$. Panel (b) in Fig.~\ref{fig:Mass} indicates that the mass ordering of $B'$ and $\eta$ is interchanged below the critical chemical potential $\mu_q^{\rm cr}$ for $m_\eta^{\rm vac}/m_\pi^{\rm vac}=1.5$ whereas such a level crossing does not take place for $m_\eta^{\rm vac}/m_\pi^{\rm vac}=1$. For this reason, the masses of $\phi_1$, $\phi_2$ and $\phi_3$ are connected to
\begin{eqnarray}
(m_{\phi_1}, m_{\phi_2},m_{\phi_3}) \to ( m_{\bar{B}'}, m_{B'}, m_{\eta} ) \ \ &&{\rm for}\ \ m_\eta^{\rm vac}/m_\pi^{\rm vac}=1 \ , \nonumber\\
(m_{\phi_1}, m_{\phi_2},m_{\phi_3}) \to ( m_{\bar{B}'}, m_{\eta}, m_{B'} ) \ \ &&{\rm for}\ \ m_\eta^{\rm vac}/m_\pi^{\rm vac}=1.5\ ,
\end{eqnarray}
at $\mu_q^{\rm cr}$. These correspondences are also reflected in the renormalization constants $Z_{\phi_i}$, as depicted in Fig.~\ref{fig:Ratio}. Indeed, the figure indicates that 
in the hadronic phase
the $Z_{\phi_i}$ are reduced to
\begin{eqnarray}
(Z_{\phi_1}, Z_{\phi_2}, Z_{\phi_3}) = (0,0,1)   \ \ &&{\rm for}\ \ m_\eta^{\rm vac}/m_\pi^{\rm vac}=1\ , \nonumber\\
(Z_{\phi_1}, Z_{\phi_2}, Z_{\phi_3}) = (0,1,0)  \ \ &&{\rm for}\ \ m_\eta^{\rm vac}/m_\pi^{\rm vac}=1.5\ . \label{ZHadronPhase}
\end{eqnarray}
Thus, $\phi_3$ ($\phi_2$) state is connected to the $\eta$-meson state in this phase when $m_\eta^{\rm vac}/m_\pi^{\rm vac}=1$ ($m_\eta^{\rm vac}/m_\pi^{\rm vac}=1.5$). Meanwhile, in a limit of $\mu_q\to\infty$, both the parameter sets read $Z_{\phi_2}\to1$ while $Z_{\phi_1},Z_{\phi_3}\to0$, reflecting a fact that the state of $\eta\sim\bar{\psi}i\gamma_5\psi$ is dominated by $\phi_2$ solely at sufficiently large $\mu_q$ where $\sigma_0$ is negligible~\cite{Suenaga:2022uqn}. It should be noted that the $\phi_1$ component in the $\eta$ state is suppressed at any chemical potential.

\begin{figure}[H] 
\begin{tabular}{cc}
\begin{minipage}{0.5\hsize}
\begin{center}
    \includegraphics[width=7.2cm]{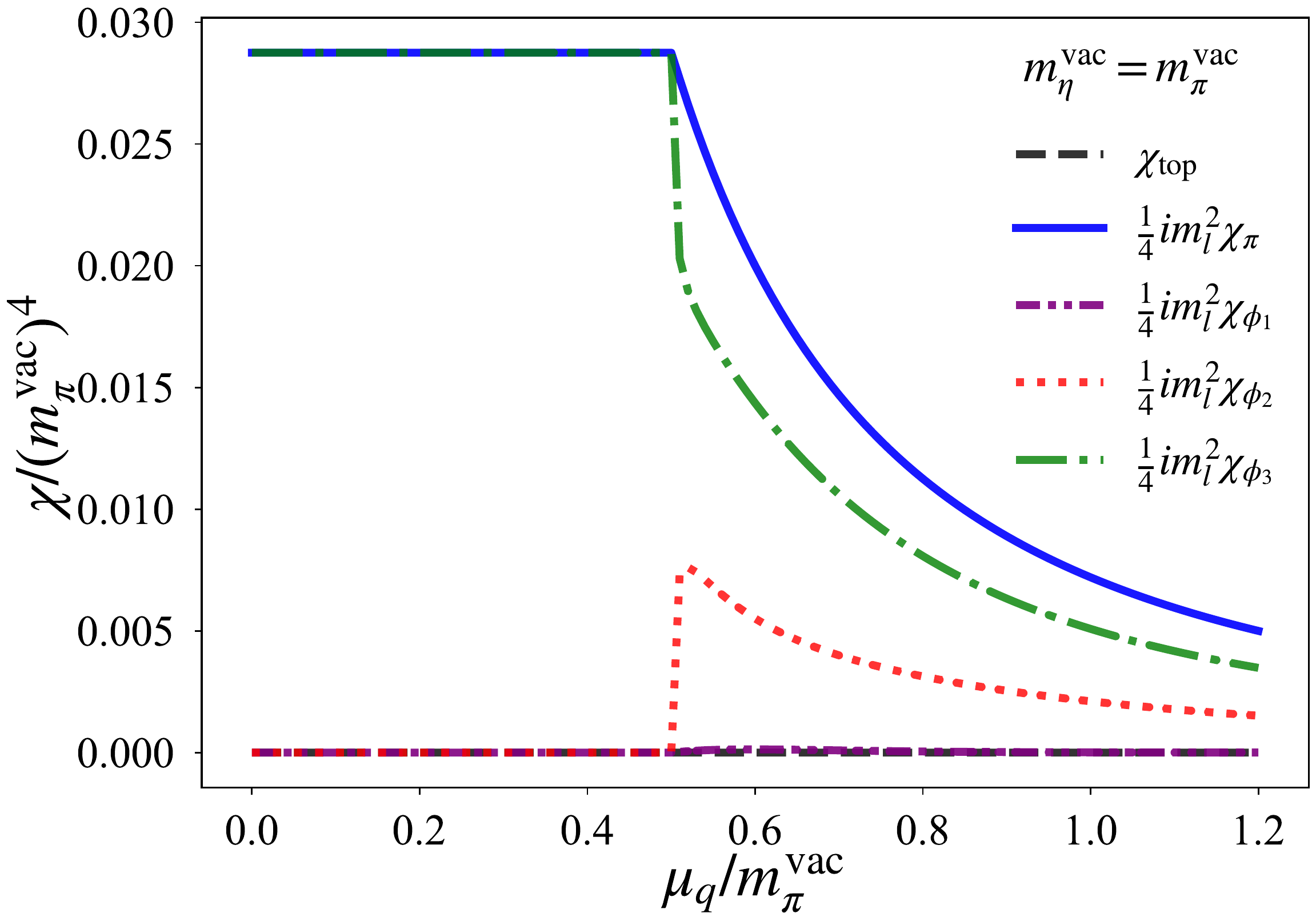}
    \subfigure{(a)}
\end{center}
\end{minipage}
\begin{minipage}{0.5\hsize}
\begin{center}
    \includegraphics[width=7.2cm]{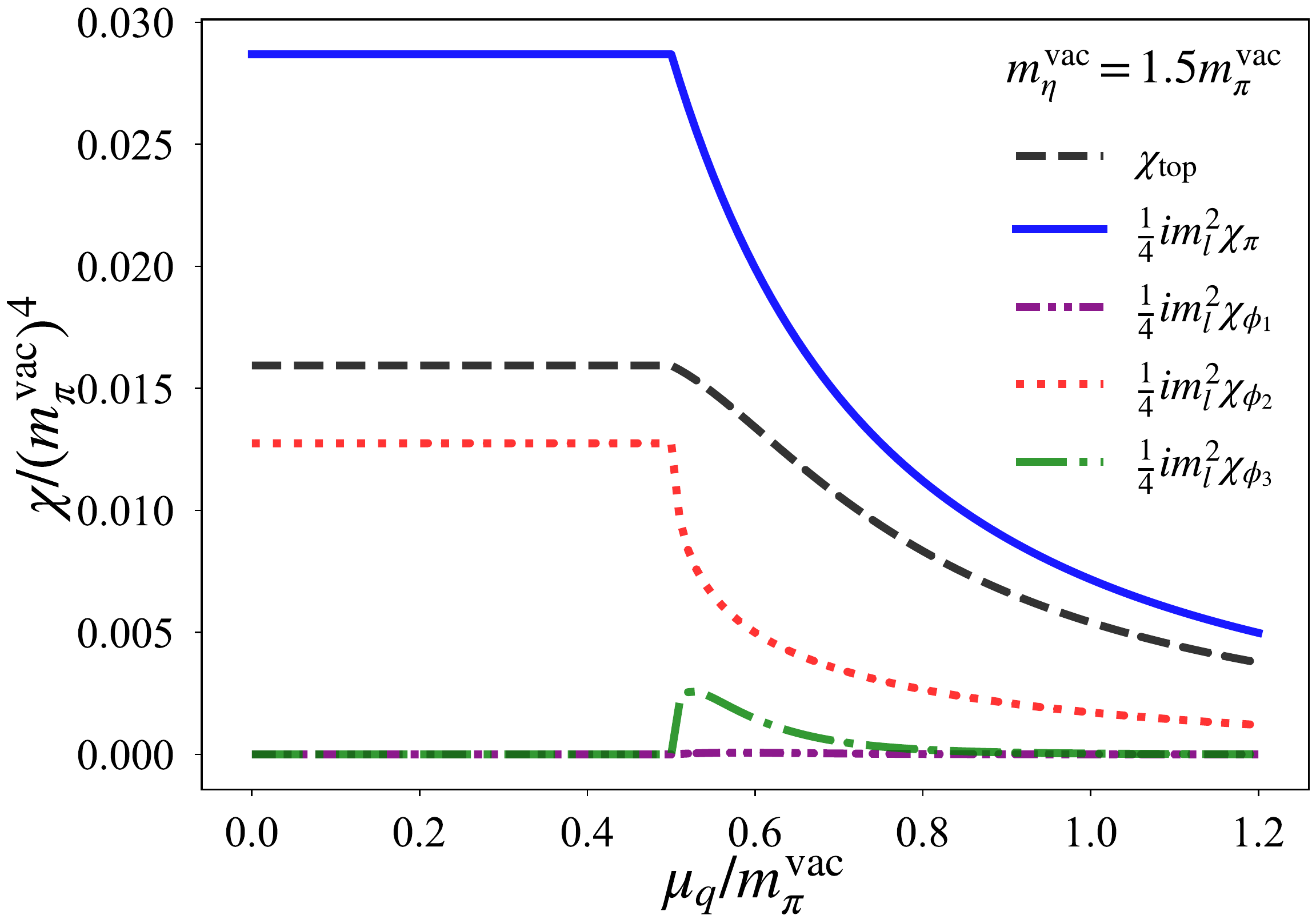}
    \subfigure{(b)}
\end{center}
\end{minipage}
\end{tabular}
\caption{
$\mu_q$ dependence of the topological susceptibility and the meson susceptibility functions for $m_\eta^{\rm vac}/m_\pi^{\rm vac}=1$ (a) and $m_\eta^{\rm vac}/m_\pi^{\rm vac}=1.5$ (b). In this figure, the susceptibilities are scaled by $(m_\pi^{\rm vac})^4$. 
}
\label{fig:ChiTop}
\end{figure}

Keeping the above properties in mind, we depict $\mu_q$ dependences of the topological susceptibility $\chi_{\rm top}$ in Fig.~\ref{fig:ChiTop}. From panel (a) one can see the topological susceptibility is always zero in the absence of the $U(1)_A$ anomaly effects. In the hadronic phase, such a trend is easily understood by a fact that $m_{\pi}^{\rm vac}$ coincides with $m_{\eta}^{\rm vac}$ together with Eq.~(\ref{chitop_meson_sus}). The null topological susceptibility in the superfluid phase is rather surprising, but it is also understood as follows. Within our linear sigma model, the KMT-type interaction is introduced to mimic the gluonic anomalous part in the non-conservation law of the $U(1)_A$ axial current in Eq.~(\ref{AxialD}). This structure is irrespective of changes of dynamical symmetry-breaking properties such as the emergence of the baryon superfluidity. 
Hence, even in the superfluid phase where the $\eta$ meson mixes with $B'_4$ and $B'_5$, 
the $\theta$ dependence in the quark mass term in Eq.~(\ref{gene_func_mass_theta}) would be rotated away under $U(1)_A$ transformation when the KMT-type interaction, i.e., the $U(1)_A$ anomaly effect, is turned off.
For this reason, the topological susceptibility defined by a second derivative with respect to $\theta$ always vanishes as long as $c=0$ is taken.

 Here, we comment on behaviors of the each contribution from $\chi_\pi$ and $\chi_{\phi_i}$ in panel (a) of Fig.~\ref{fig:ChiTop}. First, since the pion mass in the baryon superfluid phase is expressed as 
 \begin{eqnarray}
 m_\pi^2 = 4\mu_q^2\ , \label{PiMass}
 \end{eqnarray}
 the pion susceptibility function $\chi_\pi$ decreases in this phase with a power of $\mu_q^{-2}$.
 In contrast to the baryon superfluid phase, $\chi_\pi$ does not change in the hadronic phase. 
 Next, the figure indicates that, in the hadronic phase, the $\eta$-meson susceptibility function is completely dominated by $\chi_{\phi_3}$, while $\chi_{\phi_1}$ and $\chi_{\phi_2}$ vanish there. 
 Hence, $\chi_{\phi_3}$ coincides with $\chi_{\pi}$ to yield $\chi_{\rm top}=0$. This behavior is understood by panel (a) of Fig.~\ref{fig:Ratio}; the $\eta$ state in the hadronic phase is connected to the $\phi_3$ one solely. 
 Moving on to the baryon superfluid phase, we find that
 $\chi_{\phi_2}$ grows from zero and $\chi_{\phi_3}$ becomes smaller than $\chi_\pi$ to compensate the growth. Although Fig.~\ref{fig:Ratio} exhibits the significant interchange of $Z_{\phi_2}$ with $Z_{\phi_3}$ above $\mu_q\sim0.53 m_\pi^{\rm vac}$, $\chi_{\phi_2}$ is smaller than $\chi_{\phi_3}$ at any chemical potential due to the comparably strong suppression stemming from the $m_{\phi_2}^{-2}$ dependence in Eq.~(\ref{ChiEta}). Meanwhile, $\chi_{\phi_1}$ is always negligible because of the large mass suppression of $1/m_{\phi_1}^2$ and the small value of $Z_{\phi_1}$ (see in Figs.~\ref{fig:Mass} and Fig.~\ref{fig:Ratio}).  
 
The $U(1)_A$ anomaly effect represented by a nonzero $c$ in our linear sigma model can be seen from panel (b) of Fig.~\ref{fig:ChiTop}, where $m_\eta^{\rm vac}/m_\pi^{\rm vac}=1.5$ is taken. In the hadronic phase $m_\pi^{\rm vac}<m_\eta^{\rm vac}$ holds, and thus, $\chi_\eta$ becomes smaller than $\chi_\pi$, resulting the nonzero topological susceptibility. In the baryon superfluid phase, $\chi_{\rm top}$ decreases monotonically and approaches zero. The detailed analysis on this asymptotic behavior is provided in Sec.~\ref{sec:Asymptotic}. Before taking a closer look at the smooth suppression of $\chi_{\rm top}$ at larger $\mu_q$, we explain behaviors of the respective meson susceptibility functions in the presence of the KMT-type interaction. First, the $\mu_q$ dependence of $\chi_\pi$ remains the same as one without the $U(1)_A$ anomaly effects:
the $\mu_q$ scaling of $m_\pi$ in the superfluid phase,
$m_\pi^2=4\mu_q^2$, holds even when the anomaly is included.
Second, in the hadronic phase only $\chi_{\phi_2}$  contributes to the topological susceptibility while $\chi_{\phi_1}$ and $\chi_{\phi_3}$ do not, as seen from panel (b) of Fig.~\ref{fig:Ratio}. 
Third, in the superfluid phase, the finite $\chi_{\phi_3}$ is induced above $\mu_q^{\rm cr}$ owing to the bump structure of $Z_{\phi_3}$ shown in panel (b) of Fig.~\ref{fig:Ratio}.
But soon it begins to decrease and becomes negligible around $\mu_q\sim 0.8 m_\pi^{\rm vac}$ accompanied by the suppression of $Z_{\phi_3}$. Meanwhile, the abrupt suppression of $\chi_{\phi_2}$ occurs above $\mu_q^{\rm cr}$ to compensate the enhancement of $\chi_{\phi_3}$, and at larger $\mu_q$, $\chi_{\phi_2}$ gradually approaches zero in accordance with the increment of $m_{\phi_2}$. We note that $\chi_{\phi_1}$ is almost zero at any chemical potential from the same reason explained for $m_\eta^{\rm vac}/m_\pi^{\rm vac}=1$.

\subsection{Asymptotic behavior of topological susceptibility in dense baryonic matter}
\label{sec:Asymptotic}

Here, we focus on the cases for $m_\eta^{\rm vac}/m_\pi^{\rm vac}>1$, in which the finite $\chi_{\rm top}$ is provided, in order to  
delineate the asymptotic behavior of $\chi_{\rm top}$ at larger $\mu_q$.

\begin{figure}[H] 
\begin{tabular}{cc}
\begin{minipage}{1.0\hsize}
\begin{center}
    \includegraphics[width=8cm]{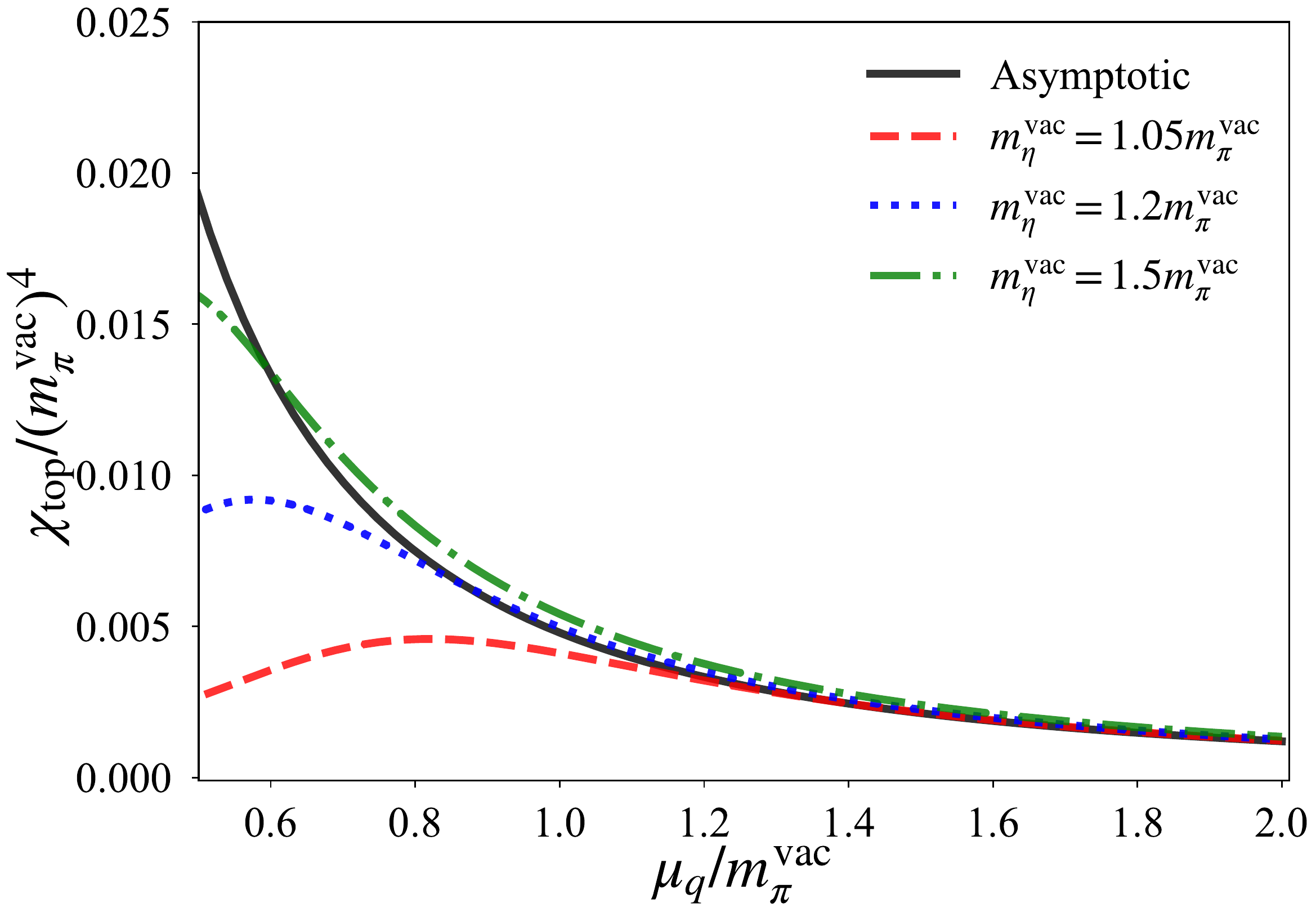}
\end{center}
\end{minipage}
\end{tabular}
\caption{Topological susceptibility in the baryon superfluid phase with several values of $m_\eta^{\rm vac}/m_\pi^{\rm vac}$.
}
\label{fig:Asymptotic}
\end{figure}

The smooth reduction of $\chi_{\rm top}$ at larger $\mu_q$ in panel (b) of Fig.~\ref{fig:ChiTop} can be explained by the continuous reduction of $\langle\bar{\psi}\psi\rangle$, as inferred from Eq.~(\ref{ChiTopQbarQ}).\footnote{A similar smooth decrease of $\chi_{\rm top}$ associated with the continuous chiral phase transition was observed in hot three-color QCD matter based on chiral model analyses~\cite{Kawaguchi:2020qvg,Kawaguchi:2020kdl,Cui:2021bqf}, and lattice simulations at physical quark masses
support such a behavior~\cite{Petreczky:2016vrs,Borsanyi:2016ksw,Bonati:2018blm}. } In order to see this behavior, we rewrite the topological susceptibility to $\chi_{\rm top} = (m_l\bar{c})^2\delta_m/(8\mu_q^2)$ with the help of Eqs.~(\ref{matched_qcon}) and~(\ref{SigmaReduction}). The dimensionless quantity $\delta_m$ is easily evaluated at sufficiently large $\mu_q$ with an assumption that $\chi_{\eta}$ is solely controlled by $\phi_2$ state. Indeed, the asymptotic behavior of $m_{\phi_2}$ is known to be $m_{\phi_2}^2\sim 12\mu_q^2$~\cite{Suenaga:2022uqn}, so that the quantity $\delta_m$ is approximated to be $\delta_m\sim 2/3$ with Eq.~(\ref{PiMass}). Therefore, the asymptotic behavior of $\chi_{\rm top}$ would be analytically fitted by
\begin{eqnarray}
\frac{\chi_{\rm top}}{(m_{\pi}^{\rm vac})^4} \sim \frac{(f_{\pi}^{\rm vac})^2}{12}\mu_q^{-2}\ ,
\label{Asymptotic}
\end{eqnarray}
where $(m_\pi^{\rm vac})^2=\sqrt{2}m_l\bar{c}/\sigma_0^{\rm vac}=m_l\bar{c}/f_\pi^{\rm vac}$ is used. 
The $\mu_q$ scaling of $\chi_{\rm top}$ coincides with that of the chiral condensate in the superfluid phase [see Eq.~(\ref{SigmaReduction})].
In Fig.~\ref{fig:Asymptotic}, we plot the topological susceptibility in the baryon superfluid phase with several values of $m_\eta^{\rm vac}/m_\pi^{\rm vac}$ with keeping input values: $m_\pi^{\rm vac}=738 {\rm MeV}$, $m_{B^\prime(\bar B^\prime)}^{\rm vac}=1611{\rm MeV}$ and $\sigma_0^{\rm vac}=250{\rm MeV}$. 
The figure shows that the asymptotic behavior of $\chi_{\rm top}$ is fitted by the analytic expression in Eq.~(\ref{Asymptotic}) well, regardless of the value of $m_\eta^{\rm vac}/m_\pi^{\rm vac}$.

 The lattice simulation performed in Ref.~\cite{Iida:2019rah} at $T= 0.45 T_c$ indicates that the topological susceptibility of QC$_2$D in the hadronic phase has a finite value with the error bars and is not influenced by the $\mu_q$ effect.
 \footnote{Here, $T_c$ denotes the pseudocritical temperature for the chiral phase transition at vanishing $\mu_q$, which are fixed to be $T_c=200$ MeV~\cite{Iida:2020emi}.} Moreover, the lattice result shows that such an approximately $\mu_q$-independent 
 behavior is further extended to the baryon superfluid phase, which obviously contradicts our model estimations. One possible scenario explaining this discrepancy is discussed in Sec.~\ref{sec:FiniteJ}. 
 In contrast to Ref.~\cite{Iida:2019rah},
 the other lattice result reported in Ref.~\cite{Astrakhantsev:2020tdl} would suggest that the topological susceptibility in the baryon superfluid phase is suppressed, as estimated by our present study.

\section{
Contamination by diquark source field
}
\label{sec:FiniteJ}

In the evaluations in Sec.~\ref{sec:Numerical}, we have included only the VEV of a scalar source field from the spurion field $\zeta_{\rm sp}$ which turns into the current quark mass as shown in Eq.~(\ref{ChiSP}). On the other hand, in lattice simulations diquark source effects incorporated from a VEV of $\zeta_B^5$ in Eq.~(\ref{Spurion}) would remain additionally, particularly in the baryon superfluid phase. Then, in this section we discuss the diquark source effects to the topological susceptibility at finite $\mu_q$ within our linear sigma model.

\subsection{
Diquark source effect on topological susceptibility
}
\label{sec:AnalyticJ}

To analytically find out contributions of the diquark source field $j$ to the topological susceptibility, we first incorporate $j$ in underlying QC$_2$D by adding the VEV of $\langle\zeta_B^5\rangle=\sqrt{2}j$ from the spurion field~(\ref{Spurion}). Now the VEV of $\zeta_{\rm sp}$ reads 
\begin{eqnarray}
\langle \zeta_{\rm sp} \rangle =
\frac{m_l}{2} E
-i\sqrt{2}j X^5 E\ .
\label{exp_sp_wj}
\end{eqnarray}
With this VEV, a diquark operator tagged with the diquark source $j$ shows up as a new ingredient in the quark mass term: 
\begin{eqnarray}
{\cal L}_{\rm QC_2D}^{\rm (mass)}
&=&
-m_l\bar \psi \psi
-j
\left(
-\frac{i}{2}\psi^TC\gamma_5\tau_c^2\tau_f^2\psi
+{\rm h.c.}
\right)\ .
\label{extraterm_wj}
\end{eqnarray}
This mass term implies that the extra term characterized by $j$ explicitly breaks the $U(1)_A$ symmetry as well as the $U(1)_B$ baryon-number symmetry. In fact, under the $U(1)_A$ axial transformation with an angle satisfying $\alpha_A=2\theta$, the generating functional of QC$_2$D with the modified mass term~(\ref{extraterm_wj}) is rotated to 
\begin{eqnarray}
Z_{\rm QC_2D}&=&
\int [d\bar{\psi} d\psi][dA]
\exp\Biggl[
i\int d^4x
\Biggl(
\bar \psi i\gamma^\mu D_\mu \psi 
-m_l\bar \psi \exp\left(i\theta/2\, \gamma_5\right)\psi \nonumber\\
&& -j
\left(
-\frac{i}{2}\psi^TC\gamma_5\tau_c^2\tau_f^2{\rm e}^{i\theta/2\gamma_5}\psi
+{\rm h.c.} \right)
-\frac{1}{4}G_{\mu\nu}^aG^{\mu\nu, a}
\Biggl)
\Biggl]\ ,
\label{gene_func_mass_theta2}
\end{eqnarray}
and hence, from Eq.~(\ref{def_chitop}) via Eq.~(\ref{VQC2D}) the net topological susceptibility $\chi_{\rm top}^{\rm w/j}$ is evaluated to be 
\begin{eqnarray}
\chi_{\rm top}^{\rm w/j} =
\chi_{\rm top}^{\rm (M)}+
\delta \chi_{\rm top}\ .
\label{chitop_wj}
\end{eqnarray}
In this expression, $\chi_{\rm top}^{\rm (M)}$ is identical to $\chi_{\rm top}$ given by Eq.~(\ref{chitop_meson_sus}), but here the superscript ``(M)'' has been attached in order to emphasize that only contributions from the meson susceptibility functions are included:
\begin{eqnarray}
\chi_{\rm top}^{\rm (M)} = \frac{i}{4}m_l^2(\chi_\pi-\chi_\eta)\ . \label{ChiMDef}
\end{eqnarray}
$\delta \chi_{\rm top}$ denotes additional contributions from $j$ which is of the form 
\begin{eqnarray}
\delta \chi_{\rm top} &=&
-\frac{1}{4}
\left[
j
\left(
-\frac{i}{2}\langle\psi^TC\gamma_5\tau_c^2\tau_f^2\psi\rangle
+{\rm h.c.}
\right)
+
ij^2 \chi_{B_5^\prime}
-2ij m_l  \chi_{ B_5^\prime\eta}
\right]\ , \label{DeltaChi}
\end{eqnarray}
where $\chi_{B_5^\prime}$ and $\chi_{B_5^\prime\eta}$ represent a susceptibility function for the $B_5^\prime$ channel
and a mixed one between the $\eta$ and $B_5^\prime$ channels, respectively. Those contributions are defined by
\begin{eqnarray}
\chi_{B_5^\prime}&=&
\int d^4x \Big\langle0\Big| 
T
\left(-\frac{1}{2}\psi^TC\tau_c^2\tau_f^2\psi+{\rm h.c.}\right)
(x) 
\left(-\frac{1}{2}\psi^TC\tau_c^2\tau_f^2\psi+{\rm h.c.}\right)
(0)
\Big|0\Big\rangle \ ,
\nonumber\\
\chi_{B_5^\prime\eta}&=&
\int d^4x \Big\langle0\Big|
T
\Big(\bar \psi i\gamma_5 \psi \Big) (x)
\left(-\frac{1}{2}\psi^TC\tau_c^2\tau_f^2\psi+{\rm h.c.}\right)
(0)
\Big|0\Big\rangle \ . \label{ChiMod}
\end{eqnarray}
The additional contributions~(\ref{DeltaChi}) can be further reduced. That is, using an identity 
\begin{eqnarray}
 -\frac{i}{2}\langle\psi^TC\gamma_5\tau_c^2\tau_f^2\psi\rangle+{\rm h.c.} =-ij\chi_{B_4}\ ,  \label{DeltaWTI}
\end{eqnarray}
with
\begin{eqnarray}
\chi_{B_4} = \int d^4x\langle0|T\left(\frac{1}{2}\psi^TC\gamma_5\tau_c^2\tau_f^2\psi +{\rm h.c.}\right)(x)\left(\frac{1}{2}\psi^TC\gamma_5\tau_c^2\tau_f^2\psi +{\rm h.c.}\right)(0)
\Big|0\Big\rangle \ ,
\end{eqnarray}
which is derived in Appendix~\ref{sec:WTIs}, the corrections $\delta\chi_{\rm top}$ in Eq.~(\ref{DeltaChi}) are rewritten in terms of the hadron susceptibility functions as
\begin{eqnarray}
\delta\chi_{\rm top} =
\chi_{\rm top}^{(\rm mix)} +
\chi_{\rm top}^{(\rm B)}\ ,
\label{DeltaChi2}
\end{eqnarray}
where\footnote{Utilizing the matching condition~(\ref{MatchDelta}) and the stationary condition for $\Delta$ in the presence of $j$, one can show that the GOR-like relation with respect to the breakdown of $U(1)_B$ baryon-number symmetry reads $f_B^2m_{B_4}^2=-j\langle\psi\psi\rangle/2$, with $f_B=\Delta/\sqrt{2}$ being the corresponding decay constant and $m_{B_4}^2=m_\pi^2-4\mu_q^2$. Here, $\langle\psi\psi\rangle$ is identical to the LHS of Eq.~(\ref{MatchDelta}). From this relation, $\chi_{\rm top}^{(\rm B)}$ can be rewritten into  
\begin{eqnarray}
\chi_{\rm top}^{(\rm B)} = \frac{f_B^2m_{B_4}^2}{2}\delta_m^{(\rm B)}\ , 
\end{eqnarray}
with $\delta_m^{(\rm B)} = 1-\chi_{B_5'}/\chi_{B_4}$, analogous to the expression for the meson sector in Eq.~(\ref{chi_top_vev_wdelta}). }
\begin{eqnarray}
\chi_{\rm top}^{(\rm mix)}= 
\frac{i}{2}m_lj\chi_{B_5'\eta}\ , \ \ \chi_{\rm top}^{(\rm B)}=
\frac{i}{4}j^2(\chi_{B_4}-\chi_{B_5'})\ .
\label{DeltaChiMeson} 
\end{eqnarray}
It is interesting to note that the baryonic contribution $\chi_{\rm top}^{(\rm B)}$ 
is proportional to the difference of $\chi_{B_4}$ and $\chi_{B_5'}$, 
which takes a partner structure similarly to $\chi_{\rm top}^{(\rm M)}$ argued in Sec.~\ref{sec:chitop_derivation}; the baryon susceptibility functions
$\chi_{B_4}$ and $\chi_{B_5'}$ are also transformed to each other under the $U(1)_B$ and $U(1)_A$ transformations. In the baryon sector, the partner structure reads
\begin{center}
\begin{tikzpicture}
  \matrix (m) [matrix of math nodes,row sep=4em,column sep=5em,minimum width=3em]
  {
     \chi_{B_4} & \chi_{B_5} \\
     \chi_{B_4^\prime} & \chi_{B_5^\prime} \\};
  \path[-stealth]
    (m-1-1) edge node [midway,left] {$U(1)_A$} (m-2-1)
            edge  node [above] {$U(1)_B$} (m-1-2)
    (m-2-1) edge node [below] {$U(1)_B$} (m-2-2)
    (m-1-2) edge node [right] {$U(1)_A$} (m-2-2)
    (m-2-1) edge node [midway,left] { } (m-1-1)
    (m-1-2) edge node [midway,left] { } (m-1-1)
    (m-2-2) edge node [midway,left] { } (m-1-2)
    (m-2-2) edge node [midway,left] { } (m-2-1);
\end{tikzpicture}
\end{center}
as explicitly derived in Appendix~\ref{sec:Trasformation}.

The susceptibility functions $\chi_{B_4}$, $\chi_{B_5'}$ and $\chi_{B_5'\eta}$ in Eq.~(\ref{DeltaChi2}) are evaluated within our linear sigma model by tracing a similar procedure in obtaining $\chi_\eta$ and $\chi_\pi$ in Sec.~\ref{sec:TSinLSM}. The $B_4$ mode does not mix with other hadrons in the low-energy limit, so $\chi_{B_4}$ is simply expressed by
\begin{eqnarray}
\chi_{B_4} = -2i\bar{c}^2\frac{1}{m_{B_4}^2}\ ,
\end{eqnarray}
where $m_{B_4}^2=m_\pi^2-4\mu_q^2$~\cite{Suenaga:2022uqn}. Meanwhile, $\chi_{B_5'}$ and $\chi_{B_5'\eta}$ are evaluated as
\begin{eqnarray}
\chi_{B_5'} = D_{B_5'B_5'}(0) \ , \ \ \chi_{B_5^\prime\eta} &=& D_{B_5^\prime\eta}(0) \ ,
\end{eqnarray}
by inverting the matrix~(\ref{DInvMatrix}), which may be expressed in terms of three contributions of $\phi_1$, $\phi_2$ and $\phi_3$ as done for $D_{\eta\eta}(0)$ in Eq.~(\ref{DEtaMix}). Based on these expressions, we numerically investigate $\mu_q$ dependences of
$\chi_{\rm top}^{\rm w/j}$ for several $j$ in the next subsection. 

It should be noted that the non-conservation law of the $U(1)_A$ current in Eq.~(\ref{AxialD}) is now modified as
\begin{eqnarray}
\partial_\mu j_A^\mu =
2m_l \bar \psi i\gamma_5 \psi
+
\left(
j\psi^TC\tau_c^2\tau_f^2\psi
+{\rm h.c.}
\right)
+\frac{g^2}{16\pi^2}\epsilon^{\mu\nu\rho\sigma}  G _{\mu\nu}^a  G^a_{\rho\sigma}\ ,
\label{ACL_wj}
\end{eqnarray}
where corrections of the diquark operator $\psi^TC\tau_c^2\tau_f^2\psi$ accompanied by the diquark source $j$ are present.

\subsection{
Diquark source effect on $\mu_q$ dependence of topological susceptibility
}
\label{sec:NumericalJ}

In the presence of the diquark source $j$, $U(1)_B$ baryon-number symmetry is explicitly broken even in the vacuum as understood from 
the modified quark mass term~ in Eq.~(\ref{extraterm_wj}). 
In other words, $\Delta$ would be always nonzero in our linear sigma model, so that the phase structures 
are expected to be modified from the ones for $j=0$.
Then, before showing numerical results of the topological susceptibility~(\ref{chitop_wj}), first we explore $\mu_q$ dependences of $\sigma_0$ and $\Delta$ corresponding to the chiral condensate and the diquark condensate, to clarify
the phase structures in the presence of $j$.

\begin{figure}[H] 
\begin{tabular}{cc}
\begin{minipage}{0.5\hsize}
\begin{center}
    \includegraphics[width=7.2cm]{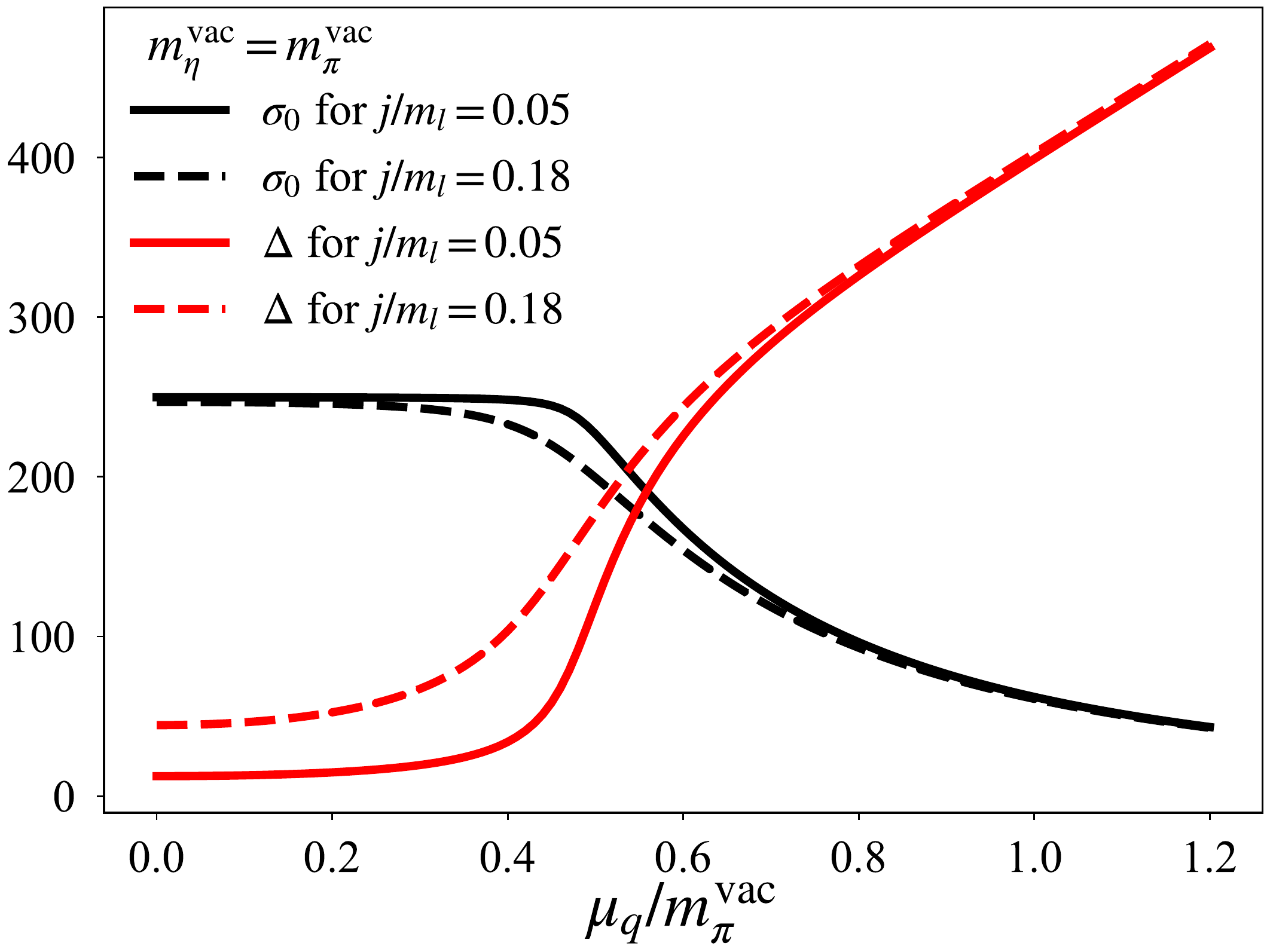}
    \subfigure{(a)}
\end{center}
\end{minipage}
\begin{minipage}{0.5\hsize}
\begin{center}
    \includegraphics[width=7.2cm]{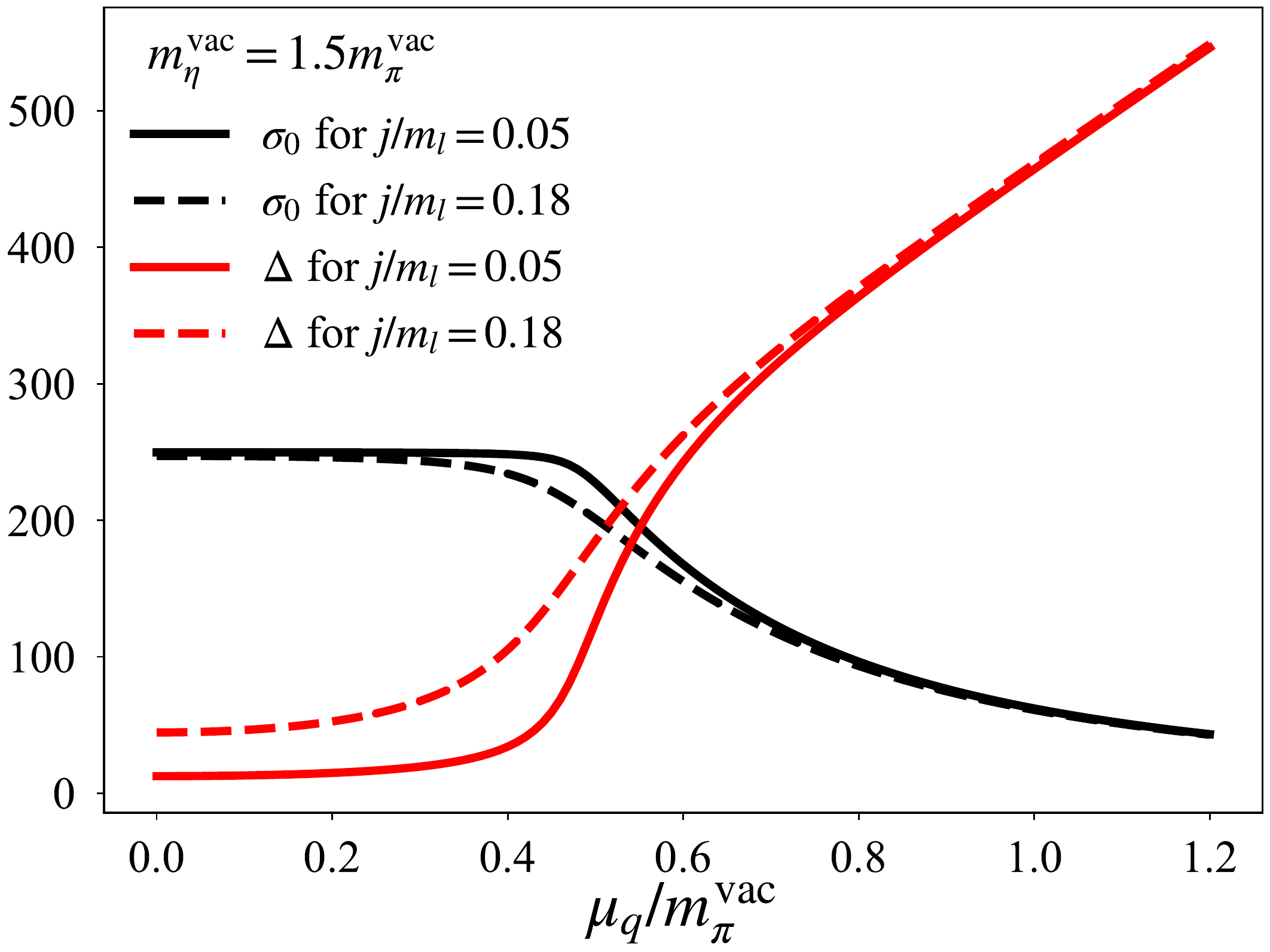}
    \subfigure{(b)}
\end{center}
\end{minipage}
\end{tabular}
\caption{
$\mu_q$ dependences of $\sigma_0$ and $\Delta$ for $j/m_l=0.05$ and $0.18$ with $m_{\eta}^{\rm vac}/m_{\pi}^{\rm vac}=1$ (a) and $m_{\eta}^{\rm vac}/m_{\pi}^{\rm vac}=1.5$ (b).
}
\label{jeff_condensates}
\end{figure}

In Fig.~\ref{jeff_condensates}, we show the $\mu_q$ dependence of the mean fields for $j/m_l=0.05$ and $0.18$ with $m_{\eta}^{\rm vac}/m_{\pi}^{\rm vac}=1$ (a) and $m_{\eta}^{\rm vac}/m_{\pi}^{\rm vac}=1.5$ (b). The figure indicates that the definite phase transition with respect to the baryon superfluidity disappears and the value of $\Delta$ continuously increases for $j\neq0$, whereas the second-order phase transition has certainly occurred for $j=0$ as seen from Fig.~\ref{expectations}. Accompanied by such a continuous change of $\Delta$, $\sigma_0$ also shows a similar smooth change. 
Besides, Fig.~\ref{jeff_condensates} indicates that $\sigma_0$ at $\mu_q\sim0$ is not significantly affected by the size of $j$ while $\Delta$ is significantly affected. This is because
the diquark source field $j$ induces 
an additional tadpole term of $\Delta$ in the effective potential in Eq.~(\ref{effpot_LSM}) which only contributes to the stationary condition of $\Delta$ directly.
Meanwhile, 
in the high-density region, $j$ contributions become negligible due to large $\mu_q$, so that the behavior of $\sigma_0$ ($\Delta$) including the $j$ effect
merges into the one for $j=0$ there.  

\begin{figure}[H] 
\begin{tabular}{cc}
\begin{minipage}{0.5\hsize}
\begin{center}
    \includegraphics[width=7.2cm]{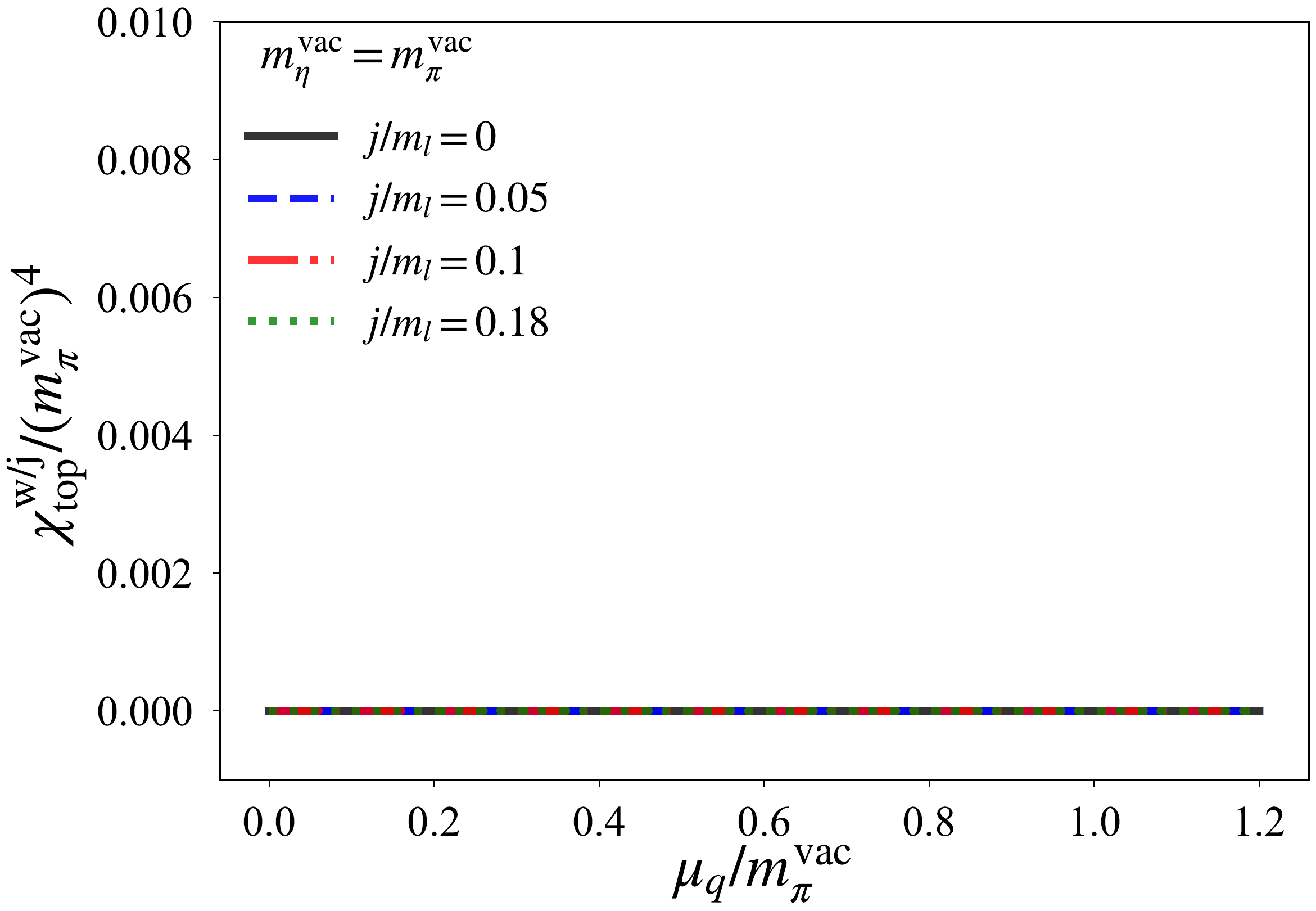}
    \subfigure{(a)}
\end{center}
\end{minipage}
\begin{minipage}{0.5\hsize}
\begin{center}
    \includegraphics[width=7.2cm]{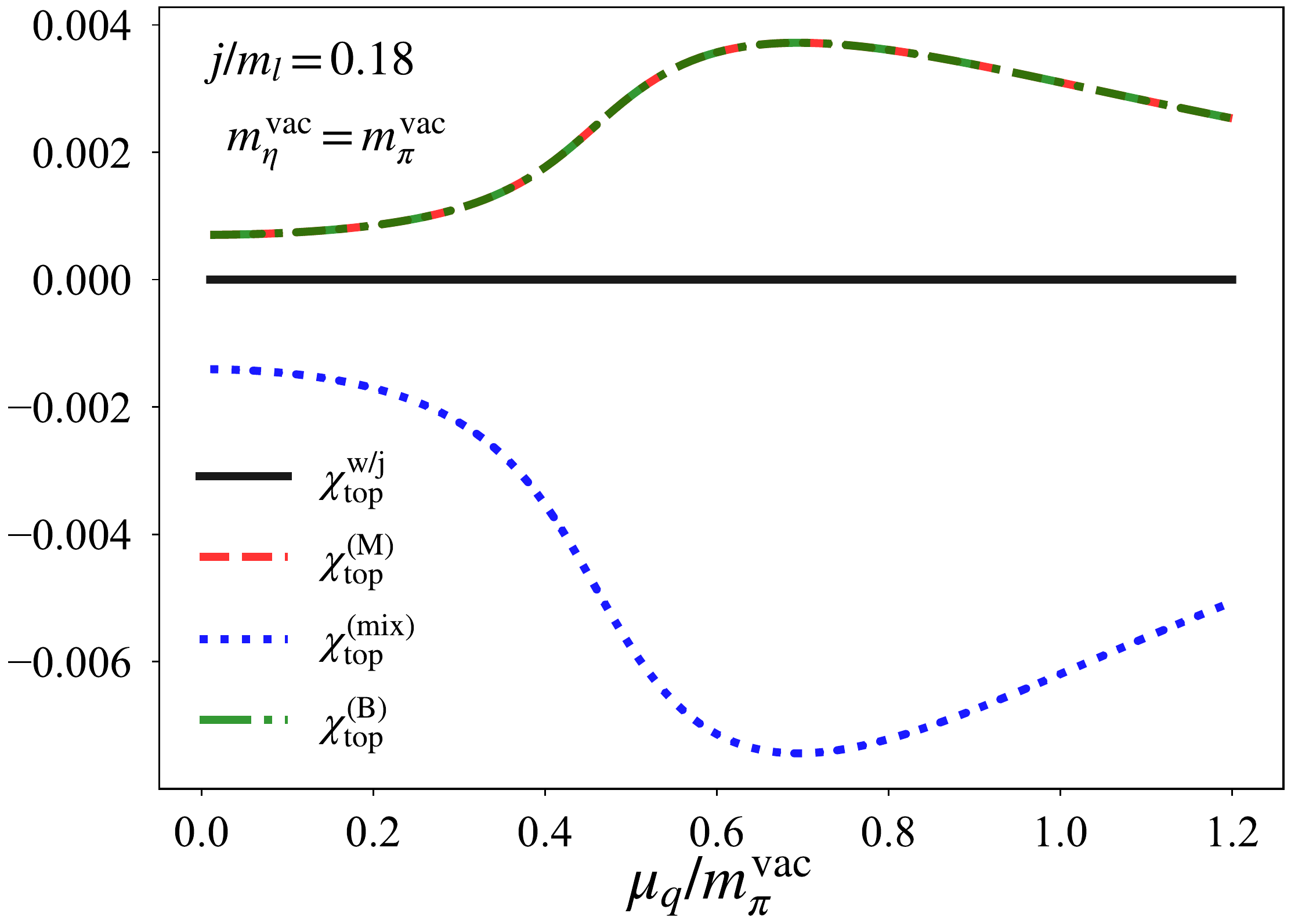}
    \subfigure{(b)}
\end{center}
\end{minipage}
\end{tabular}
\caption{$\mu_q$ dependences of the topological susceptibility including the diquark source effects in the absence of the KMT-type interaction: $m_{\eta}^{\rm vac}/m_{\pi}^{\rm vac}=1$.
}
\label{jeff_chitop_C0}
\end{figure}

\begin{figure}[H] 
\begin{tabular}{cc}
\begin{minipage}{0.5\hsize}
\begin{center}
    \includegraphics[width=7.2cm]{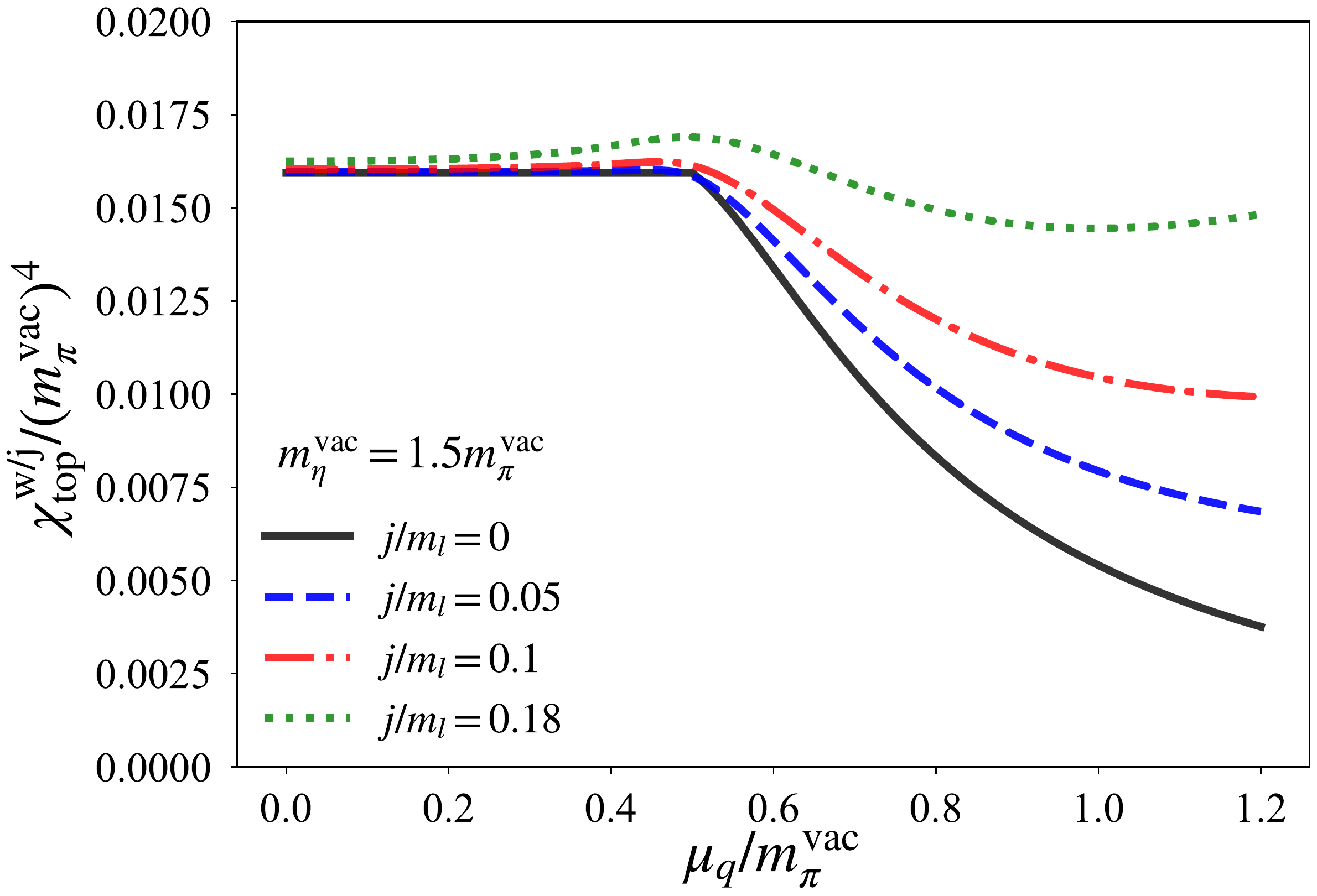}
    \subfigure{(a)}
\end{center}
\end{minipage}
\begin{minipage}{0.5\hsize}
\begin{center}
    \includegraphics[width=7.2cm]{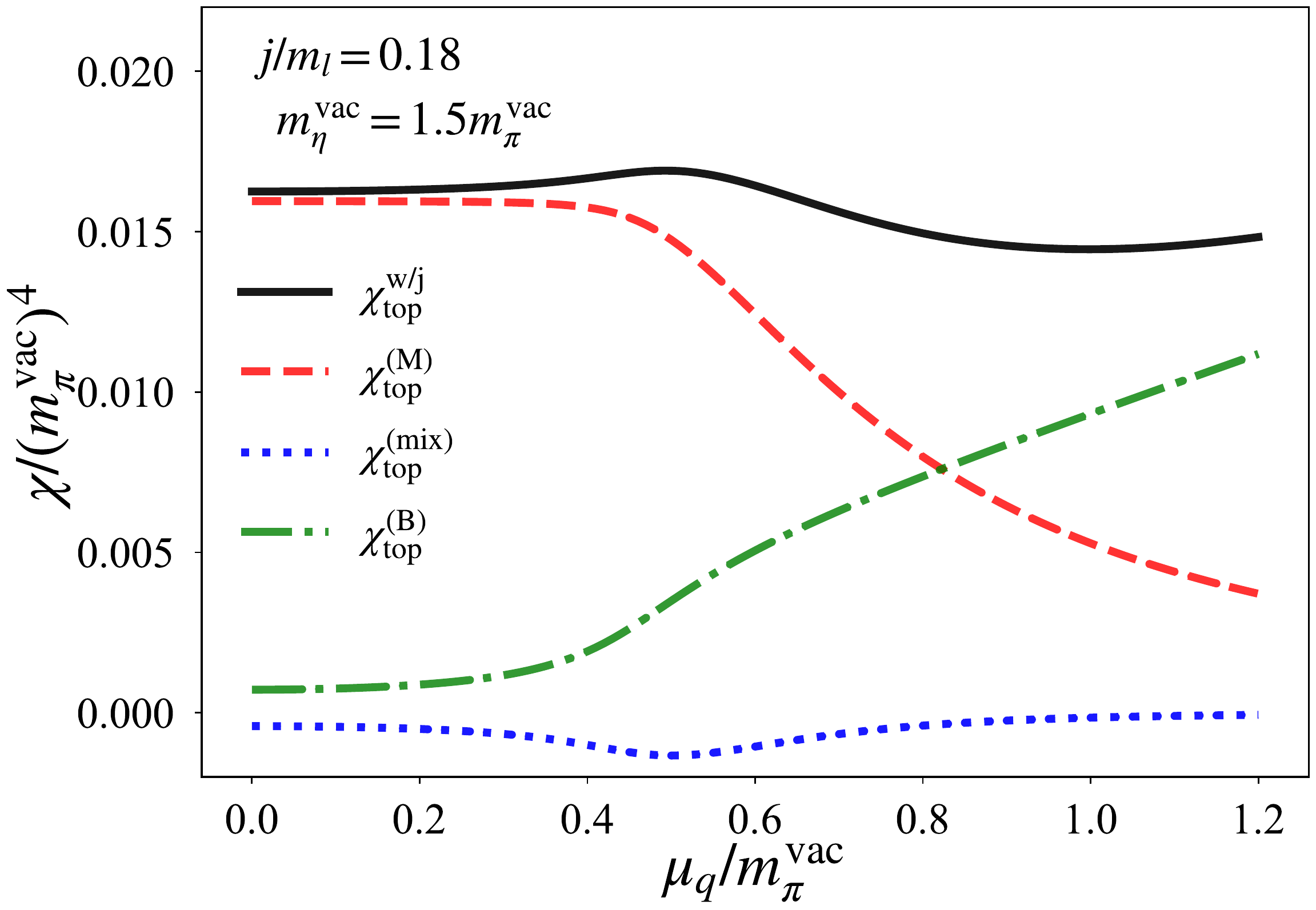}
    \subfigure{(b)}
\end{center}
\end{minipage}
\end{tabular}
\caption{
$\mu_q$ dependences of the topological susceptibility including the diquark source effects with the substantial anomaly effect of KMT-type interaction: $m_{\eta}^{\rm vac}/m_{\pi}^{\rm vac}=1.5$.
}
\label{jeff_chitop_C21}
\end{figure}

Next, we show the diquark source effects on the topological susceptibility in  Figs.~\ref{jeff_chitop_C0} and \ref{jeff_chitop_C21}.
Figure~\ref{jeff_chitop_C0} exhibits the $\mu_q$ dependence of the topological susceptibility including diquark source effects in the absence of the KMT-type interaction: $m_\eta^{\rm vac}/m_\pi^{\rm vac}=1$.
As depicted in panel (a), 
the net topological susceptibility $\chi_{\rm top}^{{\rm w/j}}$ is null at any $\mu_q$ regardless of the value of $j$. This is because the gluonic $U(1)_A$ anomaly is mimicked by only the KMT-type interaction in the linear sigma model 
even with the diquark source field $j$. 
We also analyze each component of the net topological susceptibility, $\chi_{\rm top}^{(\rm M)}$, $\chi_{\rm top}^{(\rm mix)}$ and $\chi_{\rm top}^{(\rm B)}$ defined in Eqs.~(\ref{ChiMDef}) and~(\ref{DeltaChiMeson}) for $j/m_l=0.18$, in panel (b). This panel clearly shows that the susceptibilities satisfy the relation $\chi_{\rm top}^{(\rm M)}=\chi_{\rm top}^{(\rm B)}=-\frac{1}{2}\chi_{\rm top}^{(\rm mix)}$ to result in the null $\chi_{\rm top}^{\rm w/j}$. This notable relation can be analytically derived from the anomalous WTI associated with the $U(1)_A$ transformation as shown in Appendix~\ref{sec:AotherChiTop}.

By taking the KMT-type interaction into account,
the net topological susceptibility $\chi_{\rm top}^{\rm w/j}$ becomes sensitive to the diquark source field especially above $\mu_q\approx m_{\pi}^{\rm vac}/2$ as depicted in Fig.~\ref{jeff_chitop_C21}.
Panel (a) shows that the decreasing trend of $\chi_{\rm top}^{\rm w/j}$ at higher $\mu_q$ is hindered as we take the larger value of $j$.
Notably, when taking $j/m_l=0.18$, the net topological susceptibility approximately holds the vacuum value at any $\mu_q$.
To grasp this behavior, we show separate contributions of $\chi_{\rm top}^{(\rm M)}$, $\chi_{\rm top}^{(\rm B)}$ and $\chi_{\rm top}^{(\rm mix)}$ with $j/m_l=0.18$ in panel (b) of Fig.~\ref{jeff_chitop_C21}. This figure indicates that $\chi_{\rm top}^{(\rm M)}$ is not substantially influenced by the diquark source and its decreasing behavior is governed by the smooth chiral restoration as explained in Sec.~\ref{sec:Asymptotic} in detail. In contrast, $\chi_{\rm top}^{(\rm B)}$ is enhanced above $\mu_q\approx m_{\pi}^{\rm vac}/2$, which is understood by the increment of $\Delta$. In fact, from the matching condition~(\ref{MatchDelta}) and the stationary condition for $\Delta$ in the presence of $j$, one can easily show $\chi_{\rm top}^{(\rm B)}=(j\bar{c})\Delta\delta_m^{(\rm B)}/(2\sqrt{2})$ with $\delta_m^{(\rm B)}=1-\chi_{B_5'}/\chi_{B_4}$. Here, similarly to the discussion for the meson sector in Sec.~\ref{sec:Asymptotic}, $\delta_m^{(\rm B)}$ approaches
a constant value asymptotically  in the high-density region.
Therefore, we can prove that the growth of $\chi_{\rm top}^{(\rm B)}$ can be determined by $\Delta$ at larger $\mu_q$. Hence, when the source contribution is sufficiently large, the net topological susceptibility can grow with increasing $\mu_q$. The last contribution, $\chi_{\rm top}^{(\rm mix)}$, represents a mixing susceptibility between the mesonic and baryonic sectors, and this is suppressed compared to $\chi_{\rm top}^{(\rm M)}$ and $\chi_{\rm top}^{(\rm B)}$, as long as $j$ is small, as shown in the figure. We note that, the mixing strength of $B_5'$ and $\eta$ becomes weak for larger value of $c$ as seen from $m_{B_5'\eta}^2$ in Eq.~(\ref{MEtaB5}), and hence, the larger $c$ we take, the smaller $\chi_{\rm top}^{(\rm mix)}$ we obtain.

To summarize, from the demonstration in this subsection, we have revealed that the diquark source $j$ contaminates the fate of the net topological susceptibility linked with the chiral restoration. Therefore, one can infer that 
the approximately $\mu_q$-independent behavior
of the topological susceptibility exhibited by the lattice data~\cite{Iida:2019rah} would be understood by the finite diquark source effects. 
Note that although the approximately $\mu_q$-independent behavior was found on the lattice at $T=0.45T_c$,
the temperature effects are expected to be insignificant. 
This is because
the phase structure at $T=0.45T_c$ does not significantly differ from one at $T=0$.

\section{Summary and discussion}
\label{sec:Conclusion}

In this paper, we have explored the topological susceptibility in QC$_2$D with two flavors at finite quark chemical potential $\mu_q$, to clarify
the $U(1)_A$ anomaly properties in cold and dense matter. 
With the help of the WTIs, we have found that the topological susceptibility is analytically expressed by a difference of the pion and $\eta$-meson susceptibility functions with the current quark mass. 
We have also argued that, in the low-energy regime, this expression is understood as a generalization of the one invented by Leutwyler and Smilga based on the ChPT in three-color QCD~\cite{Leutwyler:1992yt}.

In order to investigate the topological susceptibility at finite $\mu_q$, we have employed the linear sigma model in which the $U(1)_A$ anomaly effects are captured by the KMT-type determinant term, as a suitable low-energy effective theory of QC$_2$D~\cite{Suenaga:2022uqn}. This model successfully not only describes the emergence of the baryon superfluid phase but also reproduce the hadron mixings originated from the breakdown of $U(1)_B$ baryon-number symmetry there, which is indeed suggested by the lattice data~\cite{Murakami:2022lmq}. Based on a mean-field treatment, we have found that the topological susceptibility is always zero at any $\mu_q$ in both the hadronic and superfluid phases in the absence of the $U(1)_A$ anomaly effects, where the vacuum mass of pion coincides with one of $\eta$ meson. When the $U(1)_A$ anomaly effect is switched on, the nonzero and constant topological susceptibility is induced in the hadronic phase. 
Moving on to the superfluid phase, we have found that it begins to smoothly decrease  with increasing $\mu_q$. We have analytically clarified that the latter smooth decrement is fitted by $\mu_q^{-2}$ at larger $\mu_q$, reflecting the continuous restoration of chiral symmetry. This property is qualitatively the same as in hot three-color QCD matter~\cite{Petreczky:2016vrs,Borsanyi:2016ksw,Bonati:2018blm,Kawaguchi:2020qvg,Kawaguchi:2020kdl,Cui:2021bqf}. From those examinations, we can conclude that, in cold and dense QC$_2$D, roles of the topological susceptibility as an indicator for measuring the strength of $U(1)_A$ anomaly effects do not differ from those in hot three-color QCD, despite the complexity of phase structure due to the presence of the superfluidity.

In lattice simulations, effects from the diquark source would remain sizable. For this reason, we have further investigated the topological susceptibility in the presence of the diquark source. From this examination, we have revealed that the source effects enhance the topological susceptibility in accordance with the growth of the diquark condensate as $\mu_q$ increases, such that the reduction of the topological susceptibility found in the presence of the $U(1)_A$ anomaly effects can be hindered. Hence, when the source contribution is sufficiently large, the topological susceptibility can grow with increasing $\mu_q$. On the other hand, when the $U(1)_A$ anomaly effects are absent, the topological susceptibility vanishes at any value of $\mu_q$ consistently regardless of the size of the diquark source.

In closing,  
we give a list of some comments on our findings and its implications. 

\begin{itemize}
    \item
    As argued in the later part of Sec.~\ref{sec:chitop_derivation} in detail, the topological susceptibility in the vacuum is determined by only three basic observables: the pion decay constant, pion mass and $\eta$ mass, in the low-energy regime of QC$_2$D. Thus, in order to pursue a consistent understanding of $U(1)_A$ anomaly effects in low-energy QC$_2$D, we expect precise determination of both the decay constant and the $\eta$ mass as well as that of the topological susceptibility itself from lattice simulations. Those determinations would be regarded as a foundation toward more quantitative description of the topological susceptibility at finite $\mu_q$. 
    \item In the present analysis, we have used the linear sigma model based on a mean-field approach, and hence, all coupling constants in the model do not change at any $\mu_q$. On the basis of the functional renormalization group (FRG) method in three-color QCD, it was suggested that the coupling strength of the KMT-type interaction can be enhanced in medium, leading to the effective enhancement of the $U(1)_A$ anomaly effects~\cite{Fejos:2016hbp,Fejos:2017kpq,Fejos:2018dyy,Fejos:2021yod}. If this is the case, then the topological susceptibility can also be enhanced at finite $\mu_q$. Hence, analyses from effective models beyond the mean-field level such as the FRG method, in which fluctuations of the hadrons are non-perturbatively incorporated, are worth studying.
    \item 
    In our present study, we have focused on the topological susceptibility at finite $\mu_q$ but with zero temperature. Currently, the $\mu_q$ dependence on the topological susceptibility around the critical temperature has been also evaluated in the lattice QCD~\cite{Iida:2019rah}. Thus, it would be worth investigating finite temperature effects to the topological susceptibility to fit the lattice data, to pursue more comprehensive description of the $U(1)_A$ anomaly effects on the phase diagram of QC$_2$D.
    \item In this study, we have clarified that the asymptotic behavior of the topological susceptibility at larger $\mu_q$ is mostly determined by the smooth reduction of the chiral condenesate, despite the presence of the diquark condensate in dense QC$_2$D. This structure is essentially understood by a fact that the WTIs used to express the topological susceptibility in terms of the meson susceptibility functions are not altered by the diquak condensate, unless the diquark source contributions remain finite. In ordinary three-color QCD, on the other hand, the WTI associated with the pion would be modified in the color-flavor locking (CFL) phase, since the CFL configuration changes the chiral-symmetry breaking pattern due to correlations from $SU(3)_c$ color symmetry~\cite{Alford:2007xm}. For this reason, it is not clear whether a similar asymptotic behavior is derived in cold and dense three-color QCD matter, and we leave this issue for a future study. Meanwhile, the two-flavor color superconductivity (2SC) is singlet under chiral symmetry, and hence, at intermediate density regime one can expect that a qualitatively similar behavior of the topological susceptibility follows even in the presence of the 2SC phase.
\end{itemize}

\section*{Acknowledgment}
This work of M.K. is supported in part by the National Natural Science Foundation of China (NSFC) Grant  Nos: 12235016,  and  the Strategic Priority Research Program of Chinese Academy of Sciences under Grant No XDB34030000.
D.S. is supported by the RIKEN special postdoctoral researcher program. The authors thank K. Iida and E. Itou for fruitful discussion and useful information on their lattice results of the topological susceptibility.

\appendix

\section{Partner structures of the susceptibility functions}
\label{sec:Trasformation}

In this appendix, we derive partner structures of the meson and diquark-baryon susceptibility functions with respect to appropriate transformations. 

Here, we define the following composite operators
\begin{eqnarray}
&& {\cal O}_\sigma \equiv \bar{\psi}\psi\ , \ \ {\cal O}^a_{a_0} \equiv \bar{\psi}\tau_f^a\psi\ , \ \  {\cal O}_\eta \equiv \bar{\psi}i\gamma_5\psi\ , \ \ {\cal O}_\pi^a \equiv \bar{\psi}i\gamma_5\tau_f^a\psi\  , \nonumber\\
&& {\cal O}_{B_4} \equiv \frac{1}{2}\psi^TC\gamma_5\tau_c^2\tau_f^2\psi +{\rm h.c.}\ , \ \  {\cal O}_{B_5} \equiv -\frac{i}{2}\psi^TC\gamma_5\tau_c^2\tau_f^2\psi
+{\rm h.c.}\ , \nonumber\\
&& {\cal O}_{B_4'} = -\frac{i}{2}\psi^TC\tau_c^2\tau_f^2\psi-\frac{i}{2}\psi^\dagger C\tau_c^2\tau_f^2\psi^* \ , \ \ {\cal O}_{B_5'} = -\frac{1}{2}\psi^TC\tau_c^2\tau_f^2\psi + \frac{1}{2}\psi^\dagger C\tau_c^2\tau_f^2\psi^* \ . \nonumber\\
\label{InterpolatingFields}
\end{eqnarray}
The $U(1)_B$, $SU(2)_V$, $U(1)_A$ and $SU(2)_A$ rotations are generated by ($a=1,2,3$)
\begin{eqnarray}
\psi \overset{U(1)_B}{\to} {\rm e}^{-i\epsilon_B}\psi\ , \ \ \psi \overset{SU(2)_V}{\to} {\rm e}^{-i\epsilon_V^aT_f^a}\psi\ , \ \ \psi \overset{U(1)_A}{\to} {\rm e}^{-i\epsilon_A\gamma_5}\psi\ , \ \ \psi \overset{SU(2)_A}{\to}{\rm e}^{-i\epsilon_A^aT_f^a}\psi\ ,
\end{eqnarray}
respectively. Then, the meson operators ${\cal O}_\sigma$, ${\cal O}^a_{a_0}$, ${\cal O}_\eta$ and ${\cal O}^a_\pi$ are invariant under the $U(1)_B$ and $SU(2)_V$ rotations, while under the infinitesimal $U(1)_A$ and $SU(2)_A$ ones they transform as
\begin{eqnarray}
&&{\cal O}_\sigma \overset{U(1)_A}{\to} {\cal O}_\sigma-2\epsilon_A{\cal O}_\eta \ ,\ \ {\cal O}^a_{a_0} \overset{U(1)_A}{\to} {\cal O}^a_{a_0}-2\epsilon_A{\cal O}^a_{\pi}\ , \nonumber\\
&& {\cal O}_\eta \overset{U(1)_A}{\to} {\cal O}_\eta + 2\epsilon_A{\cal O}_\sigma \ , \ \ {\cal O}^a_\pi \overset{U(1)_A}{\to} {\cal O}^a_\pi + 2\epsilon_A{\cal O}^a_{a_0}\ , \label{U1ATransM}
\end{eqnarray}
and
\begin{eqnarray}
&& {\cal O}_\sigma \overset{SU(2)_A}{\to} {\cal O}_\sigma-\epsilon^a_A{\cal O}^a_\pi \ ,\ \ {\cal O}^a_{a_0} \overset{SU(2)_A}{\to} {\cal O}^a_{a_0}-\epsilon_A^a{\cal O}_{\eta}\ , \nonumber\\
&& {\cal O}_\eta \overset{SU(2)_A}{\to} {\cal O}_\eta + \epsilon_A^a{\cal O}^a_{a_0} \ , \ \ {\cal O}^a_\pi \overset{SU(2)_A}{\to} {\cal O}^a_\pi + \epsilon_A^a{\cal O}_{\sigma}\ . \label{SU2ATrans}
\end{eqnarray}
Hence, one can find the following partner structure
\begin{center}
\begin{tikzpicture}
  \matrix (m) [matrix of math nodes,row sep=4em,column sep=5em,minimum width=3em]
  {
     \chi_{\pi} & \chi_{\sigma} \\
     \chi_{a_0} & \chi_{\eta} \\};
  \path[-stealth]
    (m-1-1) edge node [midway,left] {$U(1)_A$} (m-2-1)
            edge node [above] {SU(2)} (m-1-2)
    (m-2-1) edge node [below] {SU(2)} (m-2-2)
    (m-1-2) edge node [right] {$U(1)_A$} (m-2-2)
    (m-2-1) edge node [midway,left] { } (m-1-1)
    (m-1-2) edge node [midway,left] { } (m-1-1)
    (m-2-2) edge node [midway,left] { } (m-1-2)
    (m-2-2) edge node [midway,left] { } (m-2-1);
\end{tikzpicture}
\end{center}
where the susceptibility functions are defined by
\begin{eqnarray}
&& \chi_\sigma = \int d^4x\langle0|T{\cal O}_\sigma(x){\cal O}_\sigma(0)|0\rangle\ , \ \ \chi_{a_0}\delta^{ab} = \int d^4x\langle0|T{\cal O}^a_{a_0}(x){\cal O}^b_{a_0}(0)|0\rangle \ ,\nonumber\\
&& \chi_\eta = \int d^4x\langle0|T{\cal O}_\eta(x){\cal O}_\eta(0)|0\rangle\ , \ \ \chi_{\pi}\delta^{ab} = \int d^4x\langle0|T{\cal O}^a_{\pi}(x){\cal O}^b_{\pi}(0)|0\rangle \ . \label{ChiApp1}
\end{eqnarray}

Meanwhile, the diquark-baryon operators ${\cal O}_{B_4}$, ${\cal O}_{B_5}$, ${\cal O}_{B_4'}$ and ${\cal O}_{B_5'}$ are invariant under the $SU(2)_V$ and $SU(2)_A$ rotations, while under the $U(1)_B$ and $U(1)_A$ ones they transform as
\begin{eqnarray}
&& {\cal O}_{B_4} \overset{U(1)_B}{\to} {\cal O}_{B_4} + 2\epsilon_B{\cal O}_{B_5} \ ,\ \ {\cal O}_{B_5} \overset{U(1)_B}{\to} {\cal O}_{B_5}-2\epsilon_B{\cal O}_{B_4}\ , \nonumber\\
&& {\cal O}_{B_4'} \overset{U(1)_B}{\to} {\cal O}_{B_4'} + 2\epsilon_B{\cal O}_{B_5'} \ , \ \ {\cal O}_{B_5'} \overset{U(1)_B}{\to} {\cal O}_{B_5'} -2\epsilon_B{\cal O}_{B_4'} \ ,
\label{U1BTrans}
\end{eqnarray}
and
\begin{eqnarray}
&& {\cal O}_{B_4} \overset{U(1)_A}{\to} {\cal O}_{B_4} + 2\epsilon_A {\cal O}_{B_4'} \ , \ \ {\cal O}_{B_5}  \overset{U(1)_A}{\to} {\cal O}_{B_5}+ 2\epsilon_A {\cal O}_{B_5'}\ , \nonumber\\
&& {\cal O}_{B_4'}  \overset{U(1)_A}{\to} {\cal O}_{B_4'} - 2\epsilon_A {\cal O}_{B_4} \ , \ \ {\cal O}_{B_5'}   \overset{U(1)_A}{\to} {\cal O}_{B_5'} - 2\epsilon_A {\cal O}_{B_5} \ .  \label{U1ATransB}
\end{eqnarray}
Hence, similarly to the meson sector, one can find the following partner structure
\begin{center}
\begin{tikzpicture}
  \matrix (m) [matrix of math nodes,row sep=4em,column sep=5em,minimum width=3em]
  {
     \chi_{B_4} & \chi_{B_5} \\
     \chi_{B_4^\prime} & \chi_{B_5^\prime} \\};
  \path[-stealth]
    (m-1-1) edge node [midway,left] {$U(1)_A$} (m-2-1)
            edge  node [above] {$U(1)_B$} (m-1-2)
    (m-2-1) edge node [below] {$U(1)_B$} (m-2-2)
    (m-1-2) edge node [right] {$U(1)_A$} (m-2-2)
    (m-2-1) edge node [midway,left] { } (m-1-1)
    (m-1-2) edge node [midway,left] { } (m-1-1)
    (m-2-2) edge node [midway,left] { } (m-1-2)
    (m-2-2) edge node [midway,left] { } (m-2-1);
\end{tikzpicture}
\end{center}
where the suseptibility functions are defined by
\begin{eqnarray}
&& \chi_{B_4} = \int d^4x\langle0|T{\cal O}_{B_4}(x){\cal O}_{B_4}(0)|0\rangle\ , \ \ \chi_{B_5} = \int d^4x\langle0|T{\cal O}_{B_5}(x){\cal O}_{B_5}(0)|0\rangle\ , \nonumber\\
&& \chi_{B_4'} = \int d^4x\langle0|T{\cal O}_{B_4'}(x){\cal O}_{B_4'}(0)|0\rangle\ , \ \ \chi_{B_5'} = \int d^4x\langle0|T{\cal O}_{B_5'}(x){\cal O}_{B_5'}(0)|0\rangle \ . \label{ChiApp2}
\end{eqnarray}

The infinitesimal transformation laws obtained in this appendix play important roles in deriving the WTIs employed in the present paper.

\section{Derivation of WTIs~(\ref{AWI_chiral}) and~(\ref{DeltaWTI})}
\label{sec:WTIs}

In this appendix, we derive the WTIs in Eqs.~(\ref{AWI_chiral}) and~(\ref{DeltaWTI}) which allow us to rewrite the topological susceptibility in terms of only the hadron susceptibility functions.

Toward derivation of Eq.~(\ref{AWI_chiral}), we try to perform the $SU(2)_A$ rotation in the following path integral:
\begin{eqnarray}
{\cal I}^a_\pi \equiv \int [d\bar{\psi}d\psi][dA]\, {\cal O}_\pi^a(y)\, {\rm e}^{i\int d^4x{\cal L}_{\rm QC_2D}}\ , \label{Iπi}
\end{eqnarray}
where the QC$_2$D Lagrangian of interest here includes the diquark source term in addition to the mass term as
\begin{eqnarray}
{\cal L}_{\rm QC_2D} = \bar{\psi}i\Slash{D}\psi-m_l{\cal O}_\sigma -j{\cal O}_{B_5}- \frac{1}{4}G_{\mu\nu}^aG^{\mu\nu,a}
+\theta\frac{g^2}{64\pi^2}\epsilon^{\mu\nu\rho\sigma} G_{\mu\nu}^aG_{\rho\sigma}^a \ .
\end{eqnarray}
In this Lagrangian, the covariant derivative $D_\mu\psi=(\partial_\mu-i\mu_q\delta_{\mu0}-igA_\mu^aT_c^a)\psi$ describes contributions from the quark chemical potential $\mu_q$ and couplings with the gluons $A_\mu^a$, and $G_{\mu\nu}^a=\partial_\mu A_\nu^a-\partial_\nu A_\mu^a+g\epsilon^{abc}A_\mu^b A_\nu^b$ is the gluon field strength. Under the infinitesimal local $SU(2)_A$ rotation, ${\cal O}^a_\pi$ transforms as shown in Eq.~(\ref{SU2ATrans}), while the QC$_2$D Lagrangian exhibits the following transformation law:
\begin{eqnarray}
{\cal L}_{\rm QC_2D} \overset{SU(2)_A}{\to} {\cal L}_{\rm QC_2D} + \frac{1}{2}(\partial_\mu\epsilon_A^a)j_A^{\mu,a}+m_l\epsilon_A^a{\cal O}_\pi^a\ ,
\end{eqnarray}
where $j_A^{\mu, a} \equiv \bar{\psi}\gamma^\mu\gamma_5\tau_f^a\psi$ represents the axial current. Thus, under the same rotation, Eq.~(\ref{Iπi}) transforms as
\begin{eqnarray}
{\cal I}_\pi^a &\overset{SU(2)_A}{\to}& {\cal I}_\pi^a +  \int [d\bar{\psi}d\psi][dA]\Bigg\{ \epsilon_A^a(y){\cal O}_\sigma(y) \nonumber\\
&& + i\int d^4x\epsilon_A^b(x)\left[-\frac{1}{2}\partial_\mu^x j_A^{\mu,b}(x) {\cal O}_\pi^a(y)+m_l{\cal O}_\pi^b(x){\cal O}_\pi^a(y)\right]\Bigg\}\, {\rm e}^{i\int d^4x{\cal L}_{\rm QC_2D}}\ , 
\end{eqnarray}
and imposing the invariance of ${\cal I}_\pi^a$ under the $SU(2)_A$ transformation, one can obtain the following WTI
\begin{eqnarray}
\langle{\cal O}_\sigma\rangle \delta^{ab}= i\int d^4x\left[\frac{1}{2}\partial_\mu^x\langle0|Tj_A^{\mu,b}(x){\cal O}_\pi^a(y)|0\rangle -m_l\langle0|T{\cal O}_\pi^b(x){\cal O}_\pi^a(y)|0\rangle\right]\ . \label{WTIPi}
\end{eqnarray}
Here, since the QC$_2$D Lagrangian explicitly breaks $SU(2)_L\times SU(2)_R$ chiral symmetry, there is no room for massless modes coupled to the axial current $j_A^{\mu,b}$, so that the first term of the RHS in Eq.~(\ref{WTIPi}) trivially vanishes from the surface integral. Therefore, we arrive at
\begin{eqnarray}
\langle\bar{\psi}\psi\rangle = -im_l\chi_\pi\ , \label{WTIPiApp}
\end{eqnarray}
with Eq.~(\ref{ChiApp1}).

Similarly to the above derivation, the identity~(\ref{DeltaWTI}) is also derived by focusing on the $U(1)_B$ transformation of
\begin{eqnarray}
{\cal I}_{B_4} \equiv \int [d\bar{\psi}d\psi][dA]\, {\cal O}_{B_4}(y)\, {\rm e}^{i\int d^4x{\cal L}_{\rm QC_2D}}\ . \label{IB5}
\end{eqnarray}
In fact, under the infinitesimal local $U(1)_B$ rotation, ${\cal O}_{B_4}$ transforms as in Eq.~(\ref{U1BTrans}) and the QC$_2$D Lagrangian shows the following transformation law:
\begin{eqnarray}
{\cal L}_{\rm QC_2D}\overset{U(1)_B}{\to} {\cal L}_{\rm QC_2D} + (\partial_\mu\epsilon_B)j_B^\mu + 2\epsilon_B j{\cal O}_{B_4}\ ,
\end{eqnarray}
where we have defined the vector current by $j_B^\mu\equiv\bar{\psi}\gamma^\mu\psi$. Thus, the $U(1)_B$ invariance of Eq.~(\ref{IB5}) yields
\begin{eqnarray}
2\langle{\cal O}_{B_5}\rangle =  i\int d^4x\left[\partial_\mu^x\langle0|T j_B^{\mu}(x){\cal O}_{B_4}(y)|0\rangle -2j\langle0|T{\cal O}_{B_4}(x){\cal O}_{B_4}(y)|0\rangle\right]\ . 
\end{eqnarray}
Here, the first term of the RHS vanishes since no massless modes couple to the vector current $j_B^\mu$ owing to the presence of $j$ term which violates $U(1)_B$ symmetry explicitly, and thus, one can obtain 
\begin{eqnarray}
 -\frac{i}{2}\langle\psi^TC\gamma_5\tau_c^2\tau_f^2\psi\rangle+{\rm h.c.} =-ij\chi_{B_4} 
\end{eqnarray}
by defining
\begin{eqnarray}
\chi_{B_4} = \int d^4x \langle 0|T {\cal O}_{B_4}(x) {\cal O}_{B_4}(0)   |0\rangle \ , \label{WTIB4App}
\end{eqnarray}
with Eq.~(\ref{ChiApp2}).

\section{Alternative expression of the topological susceptibility}
\label{sec:AotherChiTop}

In this appendix, we present an alternative expression of the topological susceptibility $\chi_{\rm top}$.

For this purpose, here, we consider the $U(1)_A$ transformations in the following two path integrals:
\begin{eqnarray}
{\cal I}_\eta  &\equiv& \int [d\bar{\psi}d\psi][dA]\, {\cal O}_{\eta}(y)\, {\rm e}^{i\int d^4x{\cal L}_{\rm QC_2D}}\ , \nonumber\\
{\cal I}_{B_5'}  &\equiv& \int [d\bar{\psi}d\psi][dA]\, {\cal O}_{B_5'}(y)\, {\rm e}^{i\int d^4x{\cal L}_{\rm QC_2D}}\ . \label{IEtaIB5}
\end{eqnarray}
Under the infinitesimal local $U(1)_A$ rotation, ${\cal O}_\eta$ and ${\cal O}_{B_5'}$ transform as in Eqs.~(\ref{U1ATransM}) and~(\ref{U1ATransB}) while the QC$_2$D Lagrangian shows the followng transformation law:
\begin{eqnarray}
{\cal L}_{\rm QC_2D} \overset{U(1)_A}{\to}{\cal L}_{\rm QC_2D}-(\partial_\mu\epsilon_A)j_A^\mu+\epsilon_A(2m_l{\cal O}_{\eta}-2j{\cal O}_{B_5'})\ ,
\end{eqnarray}
with the flavor-singlet $U(1)_A$ axial current defined by $j_A^\mu\equiv\bar{\psi}\gamma^\mu\gamma_5\psi$. Thus, tracing a similar procedure in deriving the WTIs~(\ref{WTIPiApp}) or~(\ref{WTIB4App}), one can find
\begin{eqnarray}
&& \langle\bar{\psi}\psi\rangle= -i\left(m_l\chi_{\eta}-j\chi_{B_5'\eta}+2\chi_{Q\eta}\right)\ , \nonumber\\
&& -\frac{i}{2}\langle\psi^TC\gamma_5\tau_c^2\tau_f^2\psi\rangle+ {\rm h.c.} = i\left(m_l\chi_{ B_5'\eta}-j\chi_{B_5'}+2\chi_{QB_5'}\right)\ ,
\label{AnomalousWTIApp}
\end{eqnarray}
where we have defined mixed susceptibility functions by
\begin{eqnarray}
&& \chi_{B_5'\eta} = \int d^4x\langle0|T{\cal O}_{B_5'}(x){\cal O}_\eta(0)|0\rangle\ , \nonumber\\
&& \chi_{Q\eta} = \int d^4x\langle0|TQ(x){\cal O}_\eta(0)|0\rangle\ , \nonumber\\
&& \chi_{QB_5'} = \int d^4x\langle0|TQ(x){\cal O}_{B_5'}(0)|0\rangle\ ,
\end{eqnarray}
with the gluonic topological operator $Q=(g^2/64\pi^2)\epsilon^{\mu\nu\rho\sigma}G_{\mu\nu}^a G_{\rho\sigma}^a$. It should be noted that the $U(1)_A$ anomaly contributions have been properly incorporated when performing the $U(1)_A$ axial transformation in ${\cal I}_\eta$ and ${\cal I}_{B_5'}$ in Eq~(\ref{IEtaIB5}).

Here, using Eqs.~(\ref{WTIPiApp}) and~(\ref{WTIB4App}), the anomalous WTIs in Eq.~(\ref{AnomalousWTIApp}) are rewritten into
\begin{eqnarray}
\chi_{\rm top}^{({\rm M})} =-\frac{1}{2}\chi_{\rm top}^{({\rm mix})}+\frac{i}{2}m_l\chi_{Q\eta} \ , \ \ \chi_{\rm top}^{({\rm B})}= -\frac{1}{2}\chi_{\rm top}^{({\rm mix})}-\frac{i}{2}j\chi_{QB_5'}\ .
\end{eqnarray}
Therefore, when we suppose that the $U(1)_A$ anomaly effects can be neglected for some reason, the operator $Q$ vanishes and 
\begin{eqnarray}
\chi_{\rm top}^{({\rm M})} = \chi_{\rm top}^{({\rm B})} = -\frac{1}{2}\chi_{\rm top}^{({\rm mix})}
\end{eqnarray}
is satisfied. Indeed, this relation is consistent with the numerical results in Fig.~\ref{jeff_chitop_C0} (a) where the $U(1)_A$ anomaly effects are absent.

\end{document}